\documentclass[11pt]{article}
\pdfoutput=1

\usepackage{tabulary}
\usepackage{epsfig}
\usepackage{psfrag}
\usepackage{latexsym}
\usepackage{indentfirst}
\usepackage{fancyhdr}
\usepackage{dsfont}
\usepackage{amsmath}
\usepackage{amssymb}
\usepackage{amsfonts}
\usepackage{mathrsfs}
\usepackage{amsthm}
\usepackage{pifont}
\usepackage{dsfont}
\usepackage{multirow}
\usepackage{array}
\usepackage{chngpage}
\usepackage{longtable}
\usepackage{cite}
\usepackage{bbold}
\usepackage{color}
\usepackage{braket}
\usepackage{colordvi}
\usepackage{fancybox}
\usepackage[footnotesize]{caption2}
\usepackage{graphicx}
\usepackage[center,footnotesize,hang]{subfigure}
\usepackage{bbm}
\usepackage[colorlinks, linkcolor=red, anchorcolor=black, citecolor=green]{hyperref}
\usepackage{multirow}
\usepackage{array}
\usepackage{textcomp}
\usepackage{enumitem}
\usepackage{mathrsfs}
\newcommand{\PreserveBackslash}[1]{\let\temp=\\#1\let\\=\temp}
\newcolumntype{C}[1]{>{\PreserveBackslash\centering}p{#1}}
\newcolumntype{R}[1]{>{\PreserveBackslash\raggedleft}p{#1}}
\newcolumntype{L}[1]{>{\PreserveBackslash\raggedright}p{#1}}
\allowdisplaybreaks

\newcommand{\bq}{\begin{eqnarray}}
\newcommand{\nq}{\end{eqnarray}}
\newcommand{\rep}[1]{\ensuremath{\boldsymbol{#1}}}

\newcommand{\bl}[1]{{\color{blue}#1}}

\begin{document}

\title{
\begin{flushright}
\hfill\mbox{\small USTC-ICTS/PCFT-20-21 } \\[5mm]
\begin{minipage}{0.2\linewidth}
\normalsize
\end{minipage}
\end{flushright}
{\Large \bf Modular Invariant $A_{4}$ Models for Quarks and Leptons with Generalized CP Symmetry \\[2mm]}}
\date{}

\author{
Chang-Yuan~Yao$^{a}$\footnote{E-mail: {\tt
yaocy@nankai.edu.cn}},~~ Jun-Nan~Lu$^{b,c}$\footnote{E-mail: {\tt
hitman@mail.ustc.edu.cn}},~~ Gui-Jun~Ding$^{b,c}$\footnote{E-mail: {\tt
dinggj@ustc.edu.cn}}
\\*[20pt]
\centerline{
\begin{minipage}{\linewidth}
\begin{center}
$^a${\it \small
School of Physics, Nankai University, Tianjin 300071, China}\\[2mm]
$^b${\it \small Peng Huanwu Center for Fundamental Theory, Hefei, Anhui 230026, China} \\[2mm]
$^c${\it \small
Interdisciplinary Center for Theoretical Study and  Department of Modern Physics,\\
University of Science and Technology of China, Hefei, Anhui 230026, China}\\
\end{center}
\end{minipage}}
\\[10mm]}

\maketitle
\thispagestyle{empty}

\begin{abstract}

We perform a systematical analysis of the $A_4$ modular models with generalized CP for the masses and flavor mixing of quarks and leptons, and the most general form of the quark and lepton mass matrices is given.  The CP invariance requires all couplings real in the chosen basis and thus the vacuum expectation value of the modulus $\tau$ uniquely breaks both the modular symmetry and CP symmetry. The phenomenologically viable models with minimal number of free parameters and the results of fit are presented. We find 20 models with 7 real free parameters that can accommodate the experimental data of lepton sector. We then apply $A_4$ modular symmetry to the quark sector to explain quark masses and CKM mixing matrix, the minimal viable quark model is found to contain 10 free real parameters. Finally, we give two predictive quark-lepton unification models which use only 16 real free parameters to explain the flavor patterns of both quarks and leptons.

\end{abstract}

\section{Introduction}

The masses and mixing parameters of both quarks and leptons have been measured precisely, however the origin of the hierarchical masses and
different mixing patterns of quarks and leptons remain elusive. Although there has been intense theoretical activity in the past decades, there is still no leading candidate for a flavor theory of quarks and leptons. From the theoretical point of view, a very appealing possibility is flavor symmetry as the guiding principle. In this setup, the lepton mixing arises from the mismatched flavor symmetry breaking patterns in the charged lepton and neutrino sectors, and similarly for the quark CKM mixing matrix. See~\cite{Feruglio:2019ktm} for for the latest review.

In conventional flavor symmetry models, the scalar potential of the flavon fields has to be cleverly designed to get the correct vacuum alignment, and generally certain auxiliary symmetry is necessary to forbid dangerous operators. As a consequence, the models look rather complex. Recently the modular invariance as flavor symmetry is suggested to overcome these drawbacks~\cite{Feruglio:2017spp}. The lepton
mass matrices would appear as the combinations of Yukawa couplings
and modular forms which are holomorphic functions of the modulus $\tau$. The vacuum expectation value of $\tau$ is the unique source of flavor symmetry breaking. In modular invariant lepton models, the neutrino masses and mixing parameters can be predicted in terms of a few input parameters and thus these models have strong predictive power.

The phenomenologies of modular invariance have widely studied in the literature, and many modular invariant models for lepton masses and mixing have been constructed by using the inhomogeneous finite modular group $\Gamma_N$ for $\Gamma_{2}\cong S_{3}$~\cite{Kobayashi:2018vbk,Kobayashi:2018wkl,Kobayashi:2019rzp,Okada:2019xqk}, $\Gamma_{3}\cong A_{4}$~\cite{Feruglio:2017spp,Criado:2018thu,Kobayashi:2018vbk,Kobayashi:2018scp,deAnda:2018ecu,Okada:2018yrn,Kobayashi:2018wkl,Novichkov:2018yse,Nomura:2019jxj,Okada:2019uoy,Nomura:2019yft,Ding:2019zxk,Okada:2019mjf,Nomura:2019lnr,Kobayashi:2019xvz,Asaka:2019vev,Gui-JunDing:2019wap,Zhang:2019ngf,Nomura:2019xsb,Wang:2019xbo,Kobayashi:2019gtp,King:2020qaj,Ding:2020yen,Okada:2020rjb,Nomura:2020opk,Okada:2020brs,Hutauruk:2020xtk}, $\Gamma_{4}\cong S_{4}$~\cite{Penedo:2018nmg,Novichkov:2018ovf,deMedeirosVarzielas:2019cyj,Kobayashi:2019mna,King:2019vhv,Criado:2019tzk,Wang:2019ovr,Gui-JunDing:2019wap,Wang:2020dbp}, $\Gamma_{5}\cong A_{5}$~\cite{Novichkov:2018nkm,Ding:2019xna,Criado:2019tzk} and $\Gamma_{7}\cong PSL(2,Z_{7})$~\cite{Ding:2020msi}. The formalism of the homogeneous finite modular group $\Gamma'_N$ which is the double covering of $\Gamma_N$ have been explored in~\cite{Liu:2019khw}. The homogeneous finite modular groups $\Gamma'_3\cong T'$~\cite{Liu:2019khw,Lu:2019vgm}, $\Gamma'_4\cong S'_4$~\cite{Novichkov:2020eep,Liu:2020akv} and $\Gamma'_5\cong A'_5$~\cite{Wang:2020lxk,Yao:2020zml} have been utilized to build lepton and quark models. Moreover, the modular invariance approach has been extended to include the rational weight modular forms, then the modular group should be extended to its metaplectic covering group and the modular forms can be decomposed into irreducible multiplets of the finite metaplectic group $\widetilde{\Gamma}_N$~\cite{Liu:2020msy}. The superstring theory requires six compact space dimensions with multiple moduli, the modular invariant supersymmetric theories with single modulus has been extended to more general automorphic supersymmetric theory where several moduli can occur naturally~\cite{Ding:2020zxw,Baur:2020yjl}. The exchange of moduli between electrons and neutrinos can induce a non-standard neutrino interactions which can leads to a shift of the neutrino mass matrix~\cite{Ding:2020yen}. Thus the presence of moduli can potentially be tested in neutrino oscillation experiments~\cite{Ding:2020yen}. The predictive power of the modular invariance approach would be improved further by including the generalized CP symmetry~(gCP) which acts on the complex modulus $\tau$ as $\tau\xrightarrow[]{CP} -\tau^{*}$ up to modular transformations~\cite{Novichkov:2019sqv,Baur:2019kwi,Acharya:1995ag,Dent:2001cc,Giedt:2002ns}.
In the symmetric basis where the generators $S$ and $T$ are represented by unitary and symmetric matrices, invariance under gCP would require all coupling constants real.

It is known that $A_{4}$ is the smallest finite modular group which admits a three-dimensional irreducible representation such that the three generations of fermion fields can be embedded into a triplet. Many $A_4$ modular models have been constructed\cite{Feruglio:2017spp,Criado:2018thu,Kobayashi:2018vbk,Kobayashi:2018scp,deAnda:2018ecu,Okada:2018yrn,Kobayashi:2018wkl,Novichkov:2018yse,Nomura:2019jxj,Okada:2019uoy,Nomura:2019yft,Ding:2019zxk,Okada:2019mjf,Nomura:2019lnr,Kobayashi:2019xvz,Asaka:2019vev,Gui-JunDing:2019wap,Zhang:2019ngf,Nomura:2019xsb,Wang:2019xbo,Kobayashi:2019gtp,King:2020qaj,Ding:2020yen,Okada:2020rjb,Nomura:2020opk,Okada:2020brs,Hutauruk:2020xtk}, and most predictive models without gCP use eight independent real parameters to describe the neutrino masses, mixing angles and CP violation phases.
In this paper, we intend to perform a systematical analysis of lepton and quark models based on $\Gamma_{3}\cong A_4$ modular symmetry with gCP, and we aim at minimizing the number of free parameters. For lepton models, we find that $20$ viable models can successfully describe the experimental data of lepton masses and mixing parameters in terms of seven real parameters including the complex modulus $\tau$. In the quark sector, in order to accommodate the measured values of quark masses and CKM mixing matrix, at least 10 free parameters are necessary
and we obtain thousands of viable quark models with 10 free real parameters. By combining the lepton and quark sectors, we find that the quark-lepton unification can be achieved with a common value of the modulus $\tau$.

The paper is organized as follows. In section~\ref{sec:Modular_forms}, we give a brief review of the basic concepts of modular symmetry and modular groups, the even weight modular forms of level 3 are constructed up to weight $8$ and they are organized into irreducible multiplets of $A_{4}$. The generalized CP symmetry compatible with $A_{4}$ modular symmetry is discussed in section~\ref{sec:gCP_A4}. In section~\ref{sec:lepton_sector}, we perform a systematical classification of $A_{4}$ modular lepton models with gCP symmetry, the phenomenologically viable models with minimal number of free parameters and the numerical results of the fit are presented. In section~\ref{sec:quark_sector}, we ultlize the $A_{4}$ modular symmetry and the gCP symmetry to explain the quark masses and mixing. The complete models for quarks and leptons are presented in section~\ref{sec:unified_models}. We conclude the paper in section~\ref{sec:conclusion}.

\section{Modular symmetry and modular forms of level $N=3$}
\label{sec:Modular_forms}

In this section, we will firstly introduce some basic concepts of modular symmetry. The modular group $\overline{\Gamma}$ can be defined from a 2-dimensional special linear group $\Gamma=SL(2,\mathbb{Z})$ with integer entries and determinant equals to 1~\cite{Bruinier2008The,diamond2005first}:
\begin{equation}
SL(2,\mathbb{Z})=\left\{\left(\begin{array}{cc}a&b\\c&d\end{array}\right)\bigg|a,b,c,d\in \mathbb{Z},ad-bc=1\right\}\,.
\end{equation}
The $SL(2,\mathbb{Z})$ group acts on the complex modulus $\tau$ in the upper half complex plane $\Im\tau >0$ via fractional linear transformations,
\begin{equation}
\label{eq:modular_trans}\tau \mapsto \gamma\tau=\gamma(\tau)=\frac{a\tau+b}{c\tau+d}\,,\quad \gamma \in SL(2,\mathbb{Z})\,.
\end{equation}
It is easy to find that $\gamma $ and $-\gamma$ induce the same linear factional transformation and thus one typically defines the projective special linear group $PSL(2,\mathbb{Z})$ by the quotient group $PSL(2,\mathbb{Z})\equiv SL(2,\mathbb{Z})/\{I,-I\}$ with $I$ is the identity matrix. The modular group $\overline{\Gamma}$ is isomorphic to $PSL(2,\mathbb{Z})$ and it has two generators $S$ and $T$ satisfying~\cite{Bruinier2008The}
\begin{equation}
S^2=(ST)^3=\mathds{1}\,.
\end{equation}
The matrix forms of $S$ and $T$ are
\begin{equation}
S=\left(
\begin{array}{cc}
0 ~&~ 1\\
-1 ~&~ 0
\end{array}
\right),\quad  T=\left(
\begin{array}{cc}
1 ~&~  1\\
0 ~&~ 1
\end{array}
\right)\,.
\end{equation}
Under the actions of $S$ and $T$, the modulus $\tau$ transform as
\begin{equation}
S: \tau \mapsto -\frac{1}{\tau},\qquad T: \tau \mapsto \tau+1\,.
\end{equation}
For a positive integer $N$, the principal congruence subgroup of level $N$ is defined as
\begin{equation}
\Gamma(N)=\left\{\left(\begin{array}{cc}a&b\\c&d\end{array}\right)\in SL(2,\mathbb{Z}),~~a-1= d-1=b=c=0~({\tt mod}~N) \right\}\,,
\end{equation}
which is a normal subgroup of the special linear group $SL(2, \mathbb{Z})$. One can find that $T^{N}$ is an element of $\Gamma(N)$. The projective principal congruence subgroup is defined as
\begin{equation}
  \overline{\Gamma}(N) = \left\{ \begin{array}{ccl} \Gamma(N)/\{\pm I \} & ~~~\text{for}~~~ & N=1, 2\\
                            \Gamma(N)  & ~~~\text{for}~~~ & N\geq 3 \end{array} \right.
\end{equation}
Notice that $-I$ does not belong to $\Gamma(N)$ for $N\geq3$. The finite modular group $\Gamma_{N}$ is the quotient group $\Gamma_N\equiv\overline{\Gamma}/\overline{\Gamma}(N)$ which can be obtained from $\overline{\Gamma}(1)$ by imposing the condition $T^N=1$. Consequently the generators $S$ and $T$ of $\Gamma_N$ satisfy the relations
\begin{equation}
S^2=(ST)^3=T^{N}=\mathds{1}\,.
\end{equation}
The finite modular groups $\Gamma_N$ with $N=2$, $3$, $4$, $5$ are isomorphic to $S_3$, $A_4$, $S_4$ and $A_5$ respectively~\cite{deAdelhartToorop:2011re}.

The modular form $f(\tau)$ of weight $k$ and level $N$ is a holomorphic function of the complex modulus $\tau$ and it transforms under the action of $\overline{\Gamma}(N)$ as follows,
\begin{equation}
f(\tau) \rightarrow f\left(\gamma \tau\right)=(c\tau+d)^kf(\tau)~~~\mathrm{for}~~\forall~\gamma = \left(
\begin{array}{cc}
a  &  b  \\
c  &  d
\end{array}
\right)\in\overline{\Gamma}(N)\,.
\end{equation}
The modular forms of weight $k$ and level $N$ span a finite dimensional linear space.
The modular forms $f_{i}(\tau)$ can be arranged into some modular multiplets $Y_{\mathbf{r}}\equiv\left(f_1(\tau), f_{2}(\tau),...\right)^{T}$ which transform as certain irreducible representation $\mathbf{r}$ of the $\Gamma_N$ for even $k$~\cite{Feruglio:2017spp,Liu:2019khw} i.e.
\begin{equation}
Y_{i}^{(k)}(\gamma\tau)=(c\tau+d)^k\rho_{ij}(\gamma)Y_{j}^{(k)}(\tau)~~~\mathrm{for}~~\forall~\gamma\in\overline{\Gamma}\,,
\end{equation}
where $\gamma$ is the representative element of the coset
$\gamma\overline{\Gamma}(N)$ in $\Gamma_N$, and $\rho_{ij}(\gamma)$ is the representation matrix of $\gamma$.

\subsection{Modular forms of level $N=3$}
\label{sec:modular_form}

The linear space of modular forms
of level 3 and weight $2k$ has dimension $2k + 1$, and the modular forms of level 3 can be organized into different irreducible representation of the inhomogeneous finite modular group $\Gamma_3\cong A_4$ up to the automorphy factor~\cite{Feruglio:2017spp}. It is known that $A_4$ is the discrete symmetry group of the rotations that leave a tethraedron invariant, or the group of the even
permutations of four objects. The $A_4$ group has three one-dimensional irreducible representations denoted by $\mathbf{1}$, $\mathbf{1}'$, $\mathbf{1}''$ and a three-dimensional representation denoted by $\mathbf{3}$. The representation matrices of the generators $S$ and $T$ are
\begin{eqnarray}
\nonumber&&\quad~~~~~~\qquad\mathbf{1}:~~ S=1, ~~~~ T=1 \,,  \\
\nonumber&&\quad~~~~~~\qquad \mathbf{1}^{\prime}:~~ S=1, ~~~~ T=\omega \,,  \\
\nonumber &&\quad~~~~~~\qquad  \mathbf{1}^{\prime\prime}:~~S=1, ~~~~ T=\omega^{2} \,, \\
\label{eq:S-T-matrices}&&\mathbf{3}:~~S=\frac{1}{3}\begin{pmatrix}
    -1& 2  & 2  \\
    2  & -1  & 2 \\
    2 & 2 & -1
\end{pmatrix}, ~~
T=\begin{pmatrix}
    1 ~&~ 0 ~&~ 0 \\
    0 ~&~ \omega ~&~ 0 \\
    0 ~&~ 0 ~&~ \omega^{2}
\end{pmatrix} \,,
\end{eqnarray}
where $\omega=e^{2\pi i/3}=-1/2+i\sqrt{3}/2$.
The decomposition of the tensor product of two triplets $\rep{\alpha}$ and $\rep{\beta}$ is
\begin{eqnarray}
\nonumber &&\left(\rep{\alpha} \otimes \rep{\beta} \right)_{\mathbf{1}}=\alpha_1\beta_1+\alpha_2\beta_3+\alpha_3\beta_2\,, \\
\nonumber &&\left(\rep{\alpha} \otimes \rep{\beta} \right)_{\mathbf{1}^{\prime}}=\alpha_3\beta_3+\alpha_1\beta_2+\alpha_2\beta_1\,, \\
\nonumber &&\left(\rep{\alpha} \otimes \rep{\beta} \right)_{\mathbf{1}^{\prime\prime}}=\alpha_2\beta_2+\alpha_1\beta_3+\alpha_3\beta_1\,, \\
\nonumber &&\left(\rep{\alpha} \otimes \rep{\beta} \right)_{\mathbf{3}_S}=\left(
\begin{array}{c}
2\alpha_1\beta_1-\alpha_2\beta_3-\alpha_3\beta_2\\
2\alpha_3\beta_3-\alpha_1\beta_2-\alpha_2\beta_1\\
2\alpha_2\beta_2-\alpha_1\beta_3-\alpha_3\beta_1)
\end{array}\right)\,, \\
\label{eq:A4-decomp} &&\left(\rep{\alpha} \otimes \rep{\beta} \right)_{\mathbf{3}_A}=\left(
\begin{array}{c}
\alpha_2\beta_3-\alpha_3\beta_2 \\
\alpha_1\beta_2-\alpha_2\beta_1 \\
\alpha_3\beta_1-\alpha_1\beta_3
\end{array}\right)\,,
\end{eqnarray}
where $\left(\rep{\alpha} \otimes \rep{\beta} \right)_{\mathbf{r}}$ denotes the contraction of $\rep{\alpha}$ and $\rep{\beta}$ into the representation $\mathbf{r}$, $\mathbf{3}_S$ and $\mathbf{3}_A$ stand for the symmetric and the antisymmetric triplet combinations respectively.

The ring of the modular forms of level 3 can be generated by three linearly independent modular forms of weight 2 which are given by~\cite{Feruglio:2017spp}:
\begin{eqnarray}
Y_1(\tau) &=& \frac{i}{2\pi}\left[ \frac{\eta'(\tau/3)}{\eta(\tau/3)}  +\frac{\eta'((\tau +1)/3)}{\eta((\tau+1)/3)}
+\frac{\eta'((\tau +2)/3)}{\eta((\tau+2)/3)} - \frac{27\eta'(3\tau)}{\eta(3\tau)}  \right], \nonumber \\
Y_2(\tau) &=& \frac{-i}{\pi}\left[ \frac{\eta'(\tau/3)}{\eta(\tau/3)}  +\omega^2\frac{\eta'((\tau +1)/3)}{\eta((\tau+1)/3)}
+\omega \frac{\eta'((\tau +2)/3)}{\eta((\tau+2)/3)}  \right] ,\nonumber \\
Y_3(\tau) &=& \frac{-i}{\pi}\left[ \frac{\eta'(\tau/3)}{\eta(\tau/3)}  +\omega\frac{\eta'((\tau +1)/3)}{\eta((\tau+1)/3)}
+\omega^2 \frac{\eta'((\tau +2)/3)}{\eta((\tau+2)/3)} \right]\,,
\end{eqnarray}
where $\eta(\tau)$ is the Dedekind eta-function
\begin{equation}
\eta(\tau)=q^{1/24} \prod_{n =1}^\infty (1-q^n),~~~~  q=e^{2\pi i\tau}\,.
\end{equation}
We can arrange the three modular functions into a vector $Y^{(2)}_{\mathbf{3}}(\tau)=\left(Y_1(\tau), Y_2(\tau), Y_3(\tau)\right)^{T}$ transforming as a triplet $\mathbf{3}$ of $A_4$,
\begin{equation}
Y^{(2)}_{\mathbf{3}}(-1/\tau)=\tau^2\rho_{\mathbf{3}}(S)Y^{(2)}_{\mathbf{3}}(\tau),~~~Y^{(2)}_{\mathbf{3}}(\tau+1)=\rho_{\mathbf{3}}(T)Y^{(2)}_{\mathbf{3}}(\tau)\,,
\end{equation}
where $\rho_{\mathbf{3}}(S)$ and $\rho_{\mathbf{3}}(T)$ are the representation matrices of $S$ and $T$ in the triplet representation $\mathbf{3}$ given in Eq.~\eqref{eq:S-T-matrices}. Multiplets of higher weight modular forms can be constructed from the tensor products of $Y^{(2)}_{\mathbf{3}}$.
At weight $k=4$, we find five independent modular forms which decompose as $\mathbf{3}\oplus\mathbf{1}\oplus\mathbf{1}'$ under $A_4$,
\begin{eqnarray}
\nonumber Y^{(4)}_{\mathbf{3}}&=&\frac{1}{2}(Y^{(2)}_{\mathbf{3}}Y^{(2)}_{\mathbf{3}})_{\mathbf{3}}=
\begin{pmatrix}
Y_1^2-Y_2 Y_3\\
Y_3^2-Y_1 Y_2\\
Y_2^2-Y_1 Y_3
\end{pmatrix}\,, \\
\nonumber Y^{(4)}_{\mathbf{1}}&=&(Y^{(2)}_{\mathbf{3}}Y^{(2)}_{\mathbf{3}})_{\mathbf{1}}=Y_1^2+2 Y_2 Y_3\,, \\
Y^{(4)}_{\mathbf{1}'}&=&(Y^{(2)}_{\mathbf{3}}Y^{(2)}_{\mathbf{3}})_{\mathbf{1}'}=Y_3^2+2 Y_1 Y_2\,.
\end{eqnarray}
Similarly there are seven modular forms of weight 6, and they can be arranged into a singlet $\mathbf{1}$ and two triplets $\mathbf{3}$ of $A_4$,
\begin{eqnarray}
Y^{(6)}_{\mathbf{1}}&=&(Y^{(2)}_{\mathbf{3}}Y^{(4)}_{\mathbf{3}})_{\mathbf{1}}=Y_1^3+Y_2^3+Y_3^3-3 Y_1 Y_2 Y_3\,,\nonumber\\
Y^{(6)}_{\mathbf{3}I}&=&Y^{(2)}_{\mathbf{3}}Y^{(4)}_{\mathbf{1}}=(Y_1^2+2Y_2Y_3)\begin{pmatrix}
Y_1\\
Y_2\\
Y_3
\end{pmatrix}\,,\nonumber\\
Y^{(6)}_{\mathbf{3}II}&=&Y^{(2)}_{\mathbf{3}}Y^{(4)}_{\mathbf{1}'}=
(Y_3^2+2 Y_1Y_2)\begin{pmatrix}
Y_3\\
Y_1\\
Y_2
\end{pmatrix}\,.
\end{eqnarray}
Finally the linearly independent weight 8 modular forms of level 3 can be decomposed into three singlets $\mathbf{1}$, $\mathbf{1'}$, $\mathbf{1''}$ and two triplets $\mathbf{3}$ under $A_4$,
\begin{eqnarray}
\nonumber Y^{(8)}_{\mathbf{1}}&=&(Y^{(2)}_{\mathbf{3}}Y^{(6)}_{\mathbf{3}I})_{\mathbf{1}}=(Y_1^2+2 Y_2 Y_3)^2\,,\\
\nonumber Y^{(8)}_{\mathbf{1'}}&=&(Y^{(2)}_{\mathbf{3}}Y^{(6)}_{\mathbf{3}I})_{\mathbf{1'}}=(Y_1^2+2 Y_2 Y_3)(Y_3^2+2 Y_1 Y_2)\,, \\
\nonumber Y^{(8)}_{\mathbf{1''}}&=&(Y^{(2)}_{\mathbf{3}}Y^{(6)}_{\mathbf{3}II})_{\mathbf{1''}}=(Y_3^2+2 Y_1 Y_2)^2\,,\\
\nonumber Y^{(8)}_{\mathbf{3}I}&=&Y^{(2)}_{\mathbf{3}}Y^{(6)}_{\mathbf{1}}=(Y_1^3+Y_2^3+Y_3^3-3 Y_1 Y_2 Y_3)\begin{pmatrix}
Y_1 \\
Y_2\\
Y_3
\end{pmatrix}\,,\\
Y^{(8)}_{\mathbf{3}II}&=&(Y^{(2)}_{\mathbf{3}}Y^{(6)}_{\mathbf{3}II})_{\mathbf{3}_A}=(Y_3^2+2 Y_1Y_2)\begin{pmatrix}
Y^2_2-Y_1Y_3\\
Y^2_1-Y_2Y_3\\
Y^2_3-Y_1Y_2
\end{pmatrix}\,.
\end{eqnarray}

\begin{table}[t!]
\centering
\begin{tabular}{|c|c|}
\hline  \hline

Modular weight $k$ & Modular form $Y^{(k)}_{\mathbf{r}}$ \\ \hline

$k=2$ & $Y^{(2)}_{\mathbf{3}}$\\ \hline

$k=4$ & $Y^{(4)}_{\mathbf{1}}, Y^{(4)}_{\mathbf{1}'}, Y^{(4)}_{\mathbf{3}}$\\ \hline

$k=6$ & $Y^{(6)}_{\mathbf{1}}, Y^{(6)}_{\mathbf{3}I}, Y^{(6)}_{\mathbf{3}II}$\\ \hline

$k=8$ & $Y^{(8)}_{\mathbf{1}}, Y^{(8)}_{\mathbf{1}'}, Y^{(8)}_{\mathbf{1}''}, Y^{(8)}_{\mathbf{3}I}, Y^{(8)}_{\mathbf{3}II}$\\ \hline \hline
\end{tabular}
\caption{\label{Tab:Level3_MF}Summary of modular forms of level 3 up to weight 8, the subscript $\mathbf{r}$ denote the transformation property under $A_4$ modular symmetry. Here $Y^{(6)}_{\mathbf{3}I}$ and $Y^{(6)}_{\mathbf{3}II}$ stand for two weight 6 modular forms transforming in the representation $\mathbf{3}$ of $A_4$. Similar conventions are adopted for $Y^{(8)}_{\mathbf{3}I}$ and $Y^{(8)}_{\mathbf{3}II}$. }
\end{table}

\section{Generalized CP consistent with $A_{4}$ symmetry}
\label{sec:gCP_A4}

The modular group $SL(2, Z)$ can consistently combine with the generalized CP symmetry by introducing another new generator which is represented by~\cite{Novichkov:2019sqv}
\begin{equation}
\mathcal{CP}=\begin{pmatrix}
1  ~&  0  \\
0  ~&  -1
\end{pmatrix}\,.
\end{equation}
Thus the modular group $\Gamma\equiv SL(2, Z)$ is enhanced to $\Gamma^{*}\equiv GL(2, Z)$. Under the action of $\mathcal{CP}$, the complex modulus $\tau$ transforms as
\begin{equation}
\tau\stackrel{\mathcal{CP}}{\longrightarrow}-\tau^{*}\,.
\end{equation}
As a consequence, the action of $\Gamma^{*}$ on the upper-half complex plane is
\begin{equation}
\begin{pmatrix}
a ~&  b  \\
c  ~& d
\end{pmatrix}\in\Gamma^{*}:~~\left\{
\begin{array}{lc}
\tau\rightarrow\frac{a\tau+b}{c\tau+d}~~\text{for}~~ad-bc=1\,,\\[0.1in]
\tau\rightarrow\frac{a\tau^{*}+b}{c\tau^{*}+d}~~\text{for}~~ad-bc=-1\,.
\end{array}
\right.
\end{equation}
Considering the CP transformation, then a modular transformation $\gamma$ and subsequently the inverse CP transformation on the modulus $\tau$, we have
\begin{equation}
\tau\stackrel{\mathcal{CP}}{\longrightarrow}-\tau^{*}\stackrel{\gamma}{\longrightarrow}-\frac{a\tau^{*}+b}{c\tau^{*}+d}\stackrel{\mathcal{CP}^{-1}}{\longrightarrow}\frac{a\tau-b}{-c\tau+d}\,.
\end{equation}
Hence the consistency condition chain $\mathcal{CP}\rightarrow \gamma\rightarrow\mathcal{CP}^{-1}$ maps the modular group element $\gamma$ into another element $u(\gamma)$\footnote{A second CP transformation can possibly be defined with $u(\gamma)=\chi(\gamma)\begin{pmatrix}
a  ~& -b  \\
-c  ~&  d
\end{pmatrix}$, where $\chi(\gamma)=\pm1$ is a homomorphism or more concretely $\chi(S)=\chi(T)=-1$~\cite{Novichkov:2020eep}. In the present work, we consider the even weight modular forms and the inhomogeneous finite modular group $\Gamma_3\cong A_4$, the modular transformation $\gamma$ is identified with $-\gamma$. Thus two generalized CP symmetries are identical.},
\begin{equation}
\gamma=\begin{pmatrix}
a  ~& b  \\
c  ~&  d
\end{pmatrix}\rightarrow u(\gamma)=\begin{pmatrix}
a  ~& -b  \\
-c  ~&  d
\end{pmatrix}=\mathcal{CP}\gamma\mathcal{CP}^{-1}\,.
\end{equation}
It is straightforward to check that $u$ is an automorphism and it preserves the structure of modular group, i.e., $u(\gamma_1)u(\gamma_2)=u(\gamma_1\gamma_2)$. Moreover there is no group element $\gamma'\in\Gamma$ such that $u(\gamma)=\gamma'\gamma\gamma'^{-1}$, therefore $u$ is an outer automorphism of the modular group. It is remarkable that $u(S)=S^{-1}$ and $u(T)=T^{-1}$.

On the other hand, the CP transformation acts on the matter field and the modular form multiplets as follow,
\begin{equation}
\psi(x)\stackrel{\mathcal{CP}}{\longrightarrow} X_{\mathbf{r}}\overline{\psi}(x_{\mathcal{P}}),~~~~Y(\tau)\stackrel{\mathcal{CP}}{\longrightarrow} Y(-\tau^{*})=X_{\mathbf{r}}Y^{*}(\tau)\,,
\end{equation}
where $x=(t, \vec{x})$ and $x_{\mathcal{P}}=(t, -\vec{x})$. Applying the consistency condition chain $\mathcal{CP}\rightarrow \gamma\rightarrow\mathcal{CP}^{-1}$ to the matter field, we can obtain the constraint on the generalized CP transformation $X_{\mathbf{r}}$,
\begin{equation}
X_{\mathbf{r}}\rho^{*}_{\mathbf{r}}(\gamma)X^{-1}_{\mathbf{r}}=\rho_{\mathbf{r}}(u(\gamma)),~~~\gamma\in\Gamma\,,
\end{equation}
which has to be fulfilled for consistent implementation of the generalized CP symmetry in the context of modular symmetry. It is sufficient to impose the above consistency condition on the generators $S$ and $T$,
\begin{equation}
X_{\mathbf{r}}\rho^{*}_{\mathbf{r}}(S)X^{-1}_{\mathbf{r}}=\rho^{-1}_{\mathbf{r}}(S),~~~X_{\mathbf{r}}\rho^{*}_{\mathbf{r}}(T)X^{-1}_{\mathbf{r}}=\rho^{-1}_{\mathbf{r}}(T)\,,
\end{equation}
In the present work, we study the finite modular group $A_4$, and both $S$ and $T$ are represented by unitary and symmetric matrices in our working basis, as shown in Eq.~\eqref{eq:S-T-matrices}. Hence the CP transformation $X_{\mathbf{r}}$ is determined to take the canonical form,
\begin{equation}
X_{\mathbf{r}}=\mathbb{1}_{\mathbf{r}}\,,
\end{equation}
up to an overall phase. Furthermore, Eq.~\eqref{eq:A4-decomp} implies that the Clebsch-Gordan coefficients are real in the chosen basis, consequently all coupling constants would be real if the theory is required invariant under the CP.

\section{Lepton models based on $A_{4}$ modular symmetry with gCP}
\label{sec:lepton_sector}

In this section,  we shall perform a systematical classification of all minimal lepton models based on the $\Gamma_3\cong A_{4}$ modular symmetry with gCP. We shall formulate our models in the framework of $\mathcal{N}=1$ global supersymmetry. In the present work, the K\"ahler potential is taken to be the minimal form as in the original work~\cite{Feruglio:2017spp}. The superpotential $\mathcal{W}(\Phi_I,\tau)$ can be expanded in power series of the supermultiplets $\Phi_I$,
\begin{equation}
\mathcal{W}(\Phi_I,\tau) =\sum_n Y_{I_1...I_n}(\tau)~ \Phi_{I_1}... \Phi_{I_n}\,,
\end{equation}
where $Y_{I_1...I_n}(\tau)$ is a modular form multiplet. Under the modular transformation, the supermultiplets $\Phi_I$ and $Y_{I_1...I_n}(\tau)$ transform as
\begin{eqnarray} \nonumber
&&\tau\to \gamma\tau=\frac{a\tau+b}{c\tau+d}\,,\qquad \qquad \qquad
~~\gamma=\begin{pmatrix}
a   &  b  \\
c   &  d
\end{pmatrix}\in SL(2,\mathbb{Z})\,,\\ \nonumber
&&\Phi_I\to (c\tau+d)^{-k_I}\rho_I(\gamma)\Phi_I\,,\\ \label{eq:modularTrs_Phi}
&&Y(\tau)\to Y(\gamma\tau)=(c\tau+d)^{k_Y}\rho_{Y}(\gamma)Y(\tau)\,.
\end{eqnarray}
where $-k_I$ and $k_{Y}$ are the modular weights of $\Phi_I$ and $Y_{I_1...I_n}$ respectively. Besides, $\rho_I(\gamma)$ and $\rho_{Y}(\gamma)$ are unitary representation of the representative element $\gamma$ in $\Gamma_{N}$. The superpotential should be invariant under the modular transformation which implies
\begin{equation}
k_Y=k_{I_1}+...+k_{I_n},~\quad~ \rho_Y\otimes\rho_{I_1}\otimes\ldots\otimes\rho_{I_n}\ni\mathbf{1}\,.
\end{equation}
In the following, the Higgs doublets $H_u$ and $H_d$ are assumed to transform as trivial singlet $\mathbf{1}$ of $A_{4}$, and their modular weights $k_{H_u}$ and $k_{H_d}$ are zero.

\subsection{\label{sec:charged_lepton_sector}Charged lepton sector}

We assume that the three generations of left-handed lepton fields transform as a triplet of $A_{4}$ modular group, while the three generations of right-handed charged leptons are assumed to transform as one-dimensional representations of $A_4$ in order to accommodate the charged lepton masses, i.e.
\begin{equation}
  \label{eq:W_E}
  L\equiv (L_{1},L_{2},L_{3})^{T}\sim \mathbf{3}\,,\quad E_{i}^{c}\sim \mathbf{1}^{l_{i}}~~\text{with}~~i=1,2,3,
\end{equation}
where $l_{1,2,3}=0,1,2$ with $\mathbf{1}^{0}\equiv \mathbf{1}$, $\mathbf{1}^{1}\equiv \mathbf{1}'$, $\mathbf{1}^{2}\equiv \mathbf{1}''$. The modular weights of $L$ and $E^{c}_{1,2,3}$ are denoted as $k_{L}$ and $k_{E^{c}_{1,2,3}}$ respectively.
The modular invariance enforces the modular forms coupling to $E^{c}_{i}LH_d$ should be a triplet of $A_{4}$, and its modular weight should be $k_{L}+k_{E^{c}_{i}}$. Thus the most general superpotential for the charged lepton masses is given by
\begin{equation}
\label{eq:charged_lepton_yukawa}
  \mathcal{W}_{E}=\sum_{i=1}^{3}\sum_{a}\alpha_{i,a}E_{i}^{c}LH_{d}Y_{\mathbf{3}a}^{(k_{L}+k_{E^{c}_{i}})}(\tau)\,,
\end{equation}
where $\alpha_{i,a}$ are coupling constants, and they are constrained to be real by gCP symmetry. The subscript $a$ indicates that there may exist multiple weight $k_{L}+k_{E^{c}_{i}}$ modular forms which transform as a triplet $\mathbf{3}$ of $A_{4}$ as shown in table~\ref{Tab:Level3_MF}.
In the following, we will denote $\alpha_{1}\equiv\alpha_{e}$, $\alpha_{2}\equiv\beta_{e}$ and $\alpha_{3}\equiv\gamma_{e}$.
From Eq.~\eqref{eq:charged_lepton_yukawa} we can see that the charged lepton mass matrix can be divided into three rows:
\begin{equation}
  M_{E}=\begin{pmatrix} R_{1}\\ R_{2}\\ R_{3} \end{pmatrix}\,,
\end{equation}
where $R_{i}$ is a $1\times 3$ sub-matrix with $i=1,2,3$. The general form of the $i-$th row of charged lepton mass matrix $R_{i}$ is determined by {modular weight $k_{L}+k_{E^{c}_{i}}$ and representation indices $l_{i}$.
Using the contraction rules of $A_{4}$ group given in Eq.~\eqref{eq:A4-decomp},
we find the general form of the $(ij)$ element of $M_{E}$ is
\begin{equation}
\label{eq:general-ele-lepton} (M_E)_{ij}=\sum_{a}\alpha_{i,a}v_{d}Y_{\mathbf{3}a,3-\text{mod}(l_{i}+j+1,3)}^{(k_{L}+k_{E^{c}_{i}})}\,.
\end{equation}
In the present work, we will be concerned with the modular forms up to weight $8$, and higher weight modular forms can be discussed in a similar way. The explicit forms of $R_{i}$ for $k_{L}+k_{E^{c}_{i}}=2,4,6,8$ are summarized in table~\ref{Tab:R_{a}}. From this table, we can see that if $k_{E^{c}_{i}}+k_{L}=2,4$, there will be a single Yukawa coupling parameter $\alpha_{i}$ for each right-handed charged lepton $E^c_i$, while there will be two Yukawa couplings $\alpha_{i,1}$ and $\alpha_{i,2}$ for $k_{E^{c}_{i}}+k_{L}=6,8$.

\begin{table}[t!]
\centering
\resizebox{1.0\textwidth}{!}{
\begin{tabular}{|c|c|c|}
\hline  \hline
  $k_{E^{c}_{i}}+k_{L}$ & Expressions of $R_{i}$ & Constraints \\
  \hline
  \multirow{4}{*}{$2,4$} & $\alpha_{i} \left(  Y^{(k_{E^{c}_{i}}+k_{L})}_{\mathbf{3},1},Y^{(k_{E^{c}_{i}}+k_{L})}_{\mathbf{3},3},Y^{(k_{E^{c}_{i}}+k_{L})}_{\mathbf{3},2}\right)v_{d}$  &  $l_{i}=0$\\
  \cline{2-3}
   & $\alpha_{i} \left(  Y^{(k_{E^{c}_{i}}+k_{L})}_{\mathbf{3},3},Y^{(k_{E^{c}_{i}}+k_{L})}_{\mathbf{3},2},Y^{(k_{E^{c}_{i}}+k_{L})}_{\mathbf{3},1}\right)v_{d}$  &  $l_{i}=1$\\
  \cline{2-3}
   & $\alpha_{i} \left(  Y^{(k_{E^{c}_{i}}+k_{L})}_{\mathbf{3},2},Y^{(k_{E^{c}_{i}}+k_{L})}_{\mathbf{3},1},Y^{(k_{E^{c}_{i}}+k_{L})}_{\mathbf{3},3}\right)v_{d}$  &  $l_{i}=2$\\
  \hline
  \multirow{4}{*}{$6,8$} & $\left(  \alpha_{i,1} Y^{(k_{E^{c}_{i}}+k_{L})}_{\mathbf{3}I,1}+\alpha_{i,2} Y^{(k_{E^{c}_{i}}+k_{L})}_{\mathbf{3}II,1},\alpha_{i,1} Y^{(k_{E^{c}_{i}}+k_{L})}_{\mathbf{3}I,3}+\alpha_{i,2} Y^{(k_{E^{c}_{i}}+k_{L})}_{\mathbf{3}II,3},\alpha_{i,1} Y^{(k_{E^{c}_{i}}+k_{L})}_{\mathbf{3}I,2}+\alpha_{i,2} Y^{(k_{E^{c}_{i}}+k_{L})}_{\mathbf{3}II,2}\right)v_{d}$  &  $l_{i}=0$\\
  \cline{2-3}
   & $\left(  \alpha_{i,1} Y^{(k_{E^{c}_{i}}+k_{L})}_{\mathbf{3}I,3}+\alpha_{i,2} Y^{(k_{E^{c}_{i}}+k_{L})}_{\mathbf{3}II,3},\alpha_{i,1} Y^{(k_{E^{c}_{i}}+k_{L})}_{\mathbf{3}I,2}+\alpha_{i,2} Y^{(k_{E^{c}_{i}}+k_{L})}_{\mathbf{3}II,2},\alpha_{i,1} Y^{(k_{E^{c}_{i}}+k_{L})}_{\mathbf{3}I,1}+\alpha_{i,2} Y^{(k_{E^{c}_{i}}+k_{L})}_{\mathbf{3}II,1}\right)v_{d}$  &  $l_{i}=1$\\
  \cline{2-3}
   & $\left(  \alpha_{i,1} Y^{(k_{E^{c}_{i}}+k_{L})}_{\mathbf{3}I,2}+\alpha_{i,2} Y^{(k_{E^{c}_{i}}+k_{L})}_{\mathbf{3}II,2},\alpha_{i,1} Y^{(k_{E^{c}_{i}}+k_{L})}_{\mathbf{3}I,1}+\alpha_{i,2} Y^{(k_{E^{c}_{i}}+k_{L})}_{\mathbf{3}II,1},\alpha_{i,1} Y^{(k_{E^{c}_{i}}+k_{L})}_{\mathbf{3}I,3}+\alpha_{i,2} Y^{(k_{E^{c}_{i}}+k_{L})}_{\mathbf{3}II,3}\right)v_{d}$  &  $l_{i}=2$\\
  \hline \hline
\end{tabular}}
\caption{\label{Tab:R_{a}}Possible structures of each row $R_{i}$ of the charged lepton mass matrix for modular forms up to weight 8, where $\alpha_{i}$, $\alpha_{i,1}$ and $\alpha_{i,2}$ are Yukawa coupling parameters.}
\end{table}

We now consider the possible structures of the charged lepton mass matrix. Combining the possible forms of $R_{i}$ with $i=1,2,3$ given in table~\ref{Tab:R_{a}}, we can directly obtain the charged lepton mass matrix.
We require that the rank of the charged lepton mass matrix not less than 3 such that any two rows of the charged lepton mass matrix can not be proportional, otherwise at least one of the changed leptons would be massless.  In other words, the three generations right-handed charged leptons must be distinguishable from each other by their modular weight and representations. Notice that the effect of exchanging the assignments of right-handed charged leptons
is to multiplying certain permutation matrices to charged lepton mass matrix from left-hand side, thus the results for the charged lepton masses and mixing matrix are not changed. Without loss of generality, we can assume $k_{E^{c}_{1}}+k_{L}\leq k_{E^{c}_{2}}+k_{L}\leq k_{E^{c}_{3}}+k_{L}$.
Thus there are the following four possible cases.

\begin{itemize}
\item{$k_{E^{c}_{1}}+k_{L}= k_{E^{c}_{2}}+k_{L}= k_{E^{c}_{3}}+k_{L}$}

In this case, all three generations of right-handed charged lepton have the same modular weights, and the values of $k_{E^{c}_{i}}+k_{L}$ can be $2,4,6,8$. To distinguish three right-handed charged leptons, their assignments of one-dimensional representations under $A_{4}$ should be different. As a result, the values of $l_{1}$, $l_{2}$ and $l_{3}$ can only be chosen as
\begin{equation}
(l_{1},l_{2},l_{3})=(0,1,2)\,,
\end{equation}
where the exchange of the values of $l_{i}$ with $i=1,2,3$ will not give any new results.
Notice that once the values of $(k_{E^{c}_{1}}+k_{L}, k_{E^{c}_{2}}+k_{L}, k_{E^{c}_{3}}+k_{L})$ and $(l_{1},l_{2},l_{3})$ are fixed, the explicit structure of charged lepton mass matrix can be read off directly from table~\ref{Tab:R_{a}}.

\item{$k_{E^{c}_{1}}+k_{L}=k_{E^{c}_{2}}+k_{L}< k_{E^{c}_{3}}+k_{L}$}

  In this case, there are 6 allowed combinations of $(k_{E^{c}_{1}}+k_{L}, k_{E^{c}_{2}}+k_{L}, k_{E^{c}_{3}}+k_{L})$, 
  \begin{equation}
    (k_{E^{c}_{1}}+k_{L}, k_{E^{c}_{2}}+k_{L}, k_{E^{c}_{3}}+k_{L}) = (2,2,4),~(2,2,6),~(2,2,8),~(4,4,6),~(4,4,8),~(6,6,8)\,.
  \end{equation}
  Since $k_{E^{c}_{1}}=k_{E^{c}_{2}}$, the values of $l_{1}$ should be distinct from $l_{2}$, while $l_{3}$ is not constrained,
  \begin{equation}
    l_{1}\neq l_{2}\,,\quad l_{1,2,3}\in \{0,1,2\}\,.
    \end{equation}
Consequently $(l_{1},l_{2},l_{3})$ can take $9$ set of values,
\begin{equation}
  (0,1,0)\,,~(0,1,1)\,,~ (0,1,2)\,,~ (0,2,0)\,,~ (0,2,1)\,,~ (0,2,2)\,,~ (1,2,0)\,,~ (1,2,1)\,,~ (1,2,2)\,.
\end{equation}

\item{$k_{E^{c}_{1}}+k_{L}< k_{E^{c}_{2}}+k_{L} = k_{E^{c}_{3}}+k_{L}$}

When the last two generations of right-handed charged leptons are assigned with the same modular weight, the values of $(k_{E^{c}_{1}}+k_{L}, k_{E^{c}_{2}}+k_{L}, k_{E^{c}_{3}}+k_{L})$ can be
\begin{equation}
    (k_{E^{c}_{1}}+k_{L}, k_{E^{c}_{2}}+k_{L}, k_{E^{c}_{3}}+k_{L}) = (2,4,4),~(2,6,6),~(2,8,8),~(4,6,6),~(4,8,8),~(6,8,8)\,.
  \end{equation}
The second and third generations of right-handed charged leptons must transform differently under $A_{4}$, therefore we have $l_{2}\neq l_{3}$ while the value of $l_{1}$ is free. Similarly we can obtain $9$ allowed assignments of $(l_{1},l_{2},l_{3})$,
\begin{equation}
  (0,0,1)\,,~(0,0,2)\,,~ (0,1,2)\,,~ (1,0,1)\,,~ (1,0,2)\,,~ (1,1,2)\,,~ (2,0,1)\,,~ (2,0,2)\,,~ (2,1,2)\,.
\end{equation}

\item{$k_{E^{c}_{1}}+k_{L}< k_{E^{c}_{2}}+k_{L} < k_{E^{c}_{3}}+k_{L}$}

If all the right-handed charged leptons have different modular weights, the values of $(k_{E^{c}_{1}}+k_{L}, k_{E^{c}_{2}}+k_{L}, k_{E^{c}_{3}}+k_{L})$ have 4 choices:
\begin{equation}
    (k_{E^{c}_{1}}+k_{L}, k_{E^{c}_{2}}+k_{L}, k_{E^{c}_{3}}+k_{L}) = (2,4,6),~(2,4,8),~(2,6,8),~(4,6,8)\,.
  \end{equation}
In this case, the representation assignments of three right-handed charged leptons are irrelevant. As a consequence, there are $27$
possible combinations of $(l_{1},l_{2},l_{3})$,
\begin{equation}
  \begin{aligned}
  &(0,0,0)\,,~(0,0,1)\,,~ (0,0,2)\,,~ (0,1,0)\,,~ (0,1,1)\,,~ (0,1,2)\,,~ (0,2,0)\,,~ (0,2,1)\,,~ (0,2,2)\,,\\
  &(1,0,0)\,,~(1,0,1)\,,~ (1,0,2)\,,~ (1,1,0)\,,~ (1,1,1)\,,~ (1,1,2)\,,~ (1,2,0)\,,~ (1,2,1)\,,~ (1,2,2)\,,\\
  &(2,0,0)\,,~(2,0,1)\,,~ (2,0,2)\,,~ (2,1,0)\,,~ (2,1,1)\,,~ (2,1,2)\,,~ (2,2,0)\,,~ (2,2,1)\,,~ (2,2,2)\,.
\end{aligned}
\end{equation}
\end{itemize}
In short, we find there are totally $4+6\times 9+6\times 9 +4\times 27=220$ possible structures of charged lepton mass matrix that will lead to non-vanishing charged lepton masses.

From table~\ref{Tab:R_{a}}, we can find that for the cases $k_{E_{i}^{c}}+k_{L}=2,4$, there is only one coupling $\alpha_{i}$ in the $i$-th row,
while there are two couplings $\alpha_{i,1}$ and $\alpha_{i,2}$ for $k_{E_{i}^{c}}+k_{L}=6,8$.
As a consequence, the number of real Yukawa coupling constants in charged lepton mass matrix is determined by the values of $k_{E_{i}^{c}}+k_{L}$ with $i=1,2,3$. If all the three modular weights $k_{E_{i}^{c}}+k_{L}$ take the values $2$ or $4$, there will be $3$ real coupling parameters in charged lepton mass matrix. If we change $k_{E_{i}^{c}}+k_{L}=2,4$ to $k_{E_{i}^{c}}+k_{L}=6,8$, one more real coupling in the $i$-th row of the mass matrix will be introduced.
Thus the minimal number of real couplings in the charged lepton mass matrix $M_{E}$ is $3$, and the maximum number is $6$. According to the number of free coupling constants in the charged lepton mass matrix, we can divide the 220 possible forms of $M_{E}$ into four distinct classes: $M_{E}^{I}$, $M_{E}^{II}$, $M_{E}^{III}$ and $M_{E}^{IV}$
in which 3, 4, 5 and 6 couplings are involved in the charged lepton mass matrix respectively.
Moreover, we find that $M_{E}^{I}$, $M_{E}^{II}$, $M_{E}^{III}$ and $M_{E}^{IV}$ include  $20$, $90$, $90$ and $20$ possible form of the charged lepton mass matrix.
The results of charged lepton mass matrix discussed above are also applicable to up and down quark mass matrices as will be shown in section~\ref{sec:quark_sector}.

\subsection{\label{sec:neutrino_sector}Neutrino sector}

We assume neutrinos to be Majorana particles and the neutrino masses are generated through the effective Weinberg operator or the type-I seesaw mechanism. If neutrino masses are described by the Weinberg operator and the three lepton doublets are assigned to an $A_4$ triplet $\mathbf{3}$, the general superpotential for neutrino masses is
\begin{eqnarray}\label{eq:Wnu_general}
  \mathcal{W}_{\nu}=\sum_{\mathbf{r}_{a}}\frac{g_{a}^{\nu}}{\Lambda}\big(H_u H_u (L L)_{\mathbf{r'}_{a}} Y_{\mathbf{r}_{a}}^{(2k_{L})}(\tau)\big)_\mathbf{1}\,,
\end{eqnarray}
where $g^{\nu}_{a}$ are coupling constants, $k_{L}$ is the modular weight of $L$, the modular weight of Higgs field $H_{u}$ is assumed to be vanishing. The modular invariance requires $\mathbf{r'}_{a}\otimes \mathbf{r}_{a}=\mathbf{1}$, and the possible assignments of $\mathbf{r}_{a}$ can be $\mathbf{1},\mathbf{1'},\mathbf{1''},\mathbf{3}$, and its value depends on the modular weight $2k_{L}$ as shown in table~\ref{Tab:Level3_MF}. The explicit forms of the elements of neutrino mass matrix can be denoted as
\begin{eqnarray}\nonumber
\label{eq:general-ele-neutrino-Wein}  (M_{\nu})_{ij}&=&\frac{v_{u}^{2}}{\Lambda}\sum_{a,b,c,d} \Big[ g^{\nu}_{a}(3\delta_{ij}-1)Y^{(2k_{L})}_{\mathbf{3}_{a},3-\text{mod}(i+j,3)}+g^{\nu}_{b}\delta_{2,\text{mod}(i+j,3)}Y^{(2k_{L})}_{\mathbf{1}_{b}}\\
&& \qquad ~~ +g^{\nu}_{c}\delta_{1,\text{mod}(i+j,3)}Y^{(2k_{L})}_{\mathbf{1'}_{c}}+g^{\nu}_{d}\delta_{0,\text{mod}(i+j,3)}Y^{(2k_{L})}_{\mathbf{1''}_{d}}\Big]\,.
\end{eqnarray}
where $i,j=1,2,3$.
If neutrino masses are generated through the type-I seesaw mechanism, for the triplet assignments of both right-handed neutrinos $N^{c}$ and left-handed lepton doublets $L$, the most general form of the superpotential in the neutrino sector is
\begin{equation}
\mathcal{W}_\nu = \sum_{\mathbf{r}_{b}} g^{D}_{b} \left(H_u (N^c L)_{\mathbf{r'}_{b}} Y_{\mathbf{r}_{b}}^{(k_{L}+k_{N^{c}})} \right)_\mathbf{1}
+ \sum_{\mathbf{r}_{c}} g^{M}_{c} \left(\Lambda (N^c N^{c})_{\mathbf{r'}_{c}} Y_{\mathbf{r}_{c}}^{(2k_{N^{c}})} \right)_\mathbf{1}\,,
\label{eq:WnuII}
\end{equation}
where $g^{D}_{b}$ and $g^{M}_{c}$ are coupling constants, $k_{N^{c}}$ is the modular weight of $N^{c}$. The modular invariance requires $\mathbf{r'}_{b}\otimes \mathbf{r}_{b}=\mathbf{1}$ and $\mathbf{r'}_{c}\otimes \mathbf{r}_{c}=\mathbf{1}$, the representation $\mathbf{r}_{b}$ and $\mathbf{r}_{c}$  can be $\mathbf{1},\mathbf{1'},\mathbf{1''},\mathbf{3}$ and they are determined by the modular weights $k_{L}$  and $k_{N^{c}}$ as shown in table~\ref{Tab:Level3_MF}.  From Eq.~\eqref{eq:WnuII}, we can read out the expressions of the elements of Dirac and Majorana neutrino mass matrices,
\begin{eqnarray}\nonumber
  (M_{D})_{ij}&=&v_{u}\sum_{a,b,c,d} \Bigg[\left( g^{D}_{a1}(3\delta_{ij}-1)+ g^{D}_{a2}(1-\delta_{ij})(-1)^{\text{mod}(i-j,3)}\right)Y^{(k_{L}+k_{N^{c}})}_{\mathbf{3}_{a},3-\text{mod}(i+j,3)}\\ \nonumber
            &&+ g^{D}_{b}\delta_{2,\text{mod}(i+j,3)}Y^{(k_{L}+k_{N^{c}})}_{\mathbf{1}_{b}}+g^{D}_{c}\delta_{1,\text{mod}(i+j,3)}Y^{(k_{L}+k_{N^{c}})}_{\mathbf{1'}_{c}}+ g^{D}_{d}\delta_{0,\text{mod}(i+j,3)}Y^{(k_{L}+k_{N^{c}})}_{\mathbf{1''}_{d}}\Bigg]\,,\\ \nonumber
 (M_{N})_{ij}&=&\Lambda \sum_{a,b,c,d} \Big[ g^{M}_{a}(3\delta_{ij}-1)Y^{(2k_{N^{c}})}_{\mathbf{3}_{a},3-\text{mod}(i+j,3)}+g^{M}_{b}\delta_{2,\text{mod}(i+j,3)}Y^{(2k_{N^{c}})}_{\mathbf{1}_{b}}\\
\label{eq:general-ele-neutrino-SS} && \qquad ~~ +g^{M}_{c}\delta_{1,\text{mod}(i+j,3)}Y^{(2k_{N^{c}})}_{\mathbf{1'}_{c}}+g^{M}_{d}\delta_{0,\text{mod}(i+j,3)}Y^{(2k_{N^{c}})}_{\mathbf{1''}_{d}}\Big]\,.
\end{eqnarray}
The effective light neutrino mass matrix in the type-I seesaw models is given by the seesaw formula,
\begin{equation}
  M_{\nu}=-M_{D}^{T}M_{N}^{-1}M_{D}\,.
\end{equation}
We are only interested in the models with less free parameters, and we list the possible neutrino models in table~\ref{tab:neutrino}, for which the resulting light neutrino mass matrices contain less than $4$ free real parameters besides the complex modulus $\tau$.

\begin{table}[t!]
\centering
\resizebox{1.0\textwidth}{!}{
\begin{tabular}{|c|c|c|} \hline\hline

  & $k_L$, $k_{N^c}$ &    Neutrino mass matrix  \\
  \hline
   &   &     \\ [-0.15in]
$W_1$ & 1, ---  & {\scriptsize $ M_\nu = g^{\nu} \begin{pmatrix}
  2Y^{(2)}_{\mathbf{3},1} ~&~ -Y^{(2)}_{\mathbf{3},3} ~&~ -Y^{(2)}_{\mathbf{3},2} \\
 -Y^{(2)}_{\mathbf{3},3} ~&~ 2Y^{(2)}_{\mathbf{3},2}  ~&~ -Y^{(2)}_{\mathbf{3},1}  \\
 -Y^{(2)}_{\mathbf{3},2} ~&~ -Y^{(2)}_{\mathbf{3},1} ~&~2Y^{(2)}_{\mathbf{3},3} \\
\end{pmatrix}\dfrac{v^2_u}{\Lambda}$}  \\
  &  &    \\ [-0.15in] \hline
   &   &     \\ [-0.15in]
$W_2$ & 2, ---  & {\scriptsize $   M_\nu = \left[g^{\nu}_{1}
\begin{pmatrix}
 2Y^{(4)}_{\mathbf{3},1} ~&~ -Y^{(4)}_{\mathbf{3},3} ~&~ -Y^{(4)}_{\mathbf{3},2} \\
-Y^{(4)}_{\mathbf{3},3} ~&~  2Y^{(4)}_{\mathbf{3},2} ~&~ -Y^{(4)}_{\mathbf{3},1}  \\
-Y^{(4)}_{\mathbf{3},2} ~&~ -Y^{(4)}_{\mathbf{3},1} ~&~2Y^{(4)}_{\mathbf{3},3}
\end{pmatrix}
+g^{\nu}_{2}Y^{(4)}_{\mathbf{1}}\begin{pmatrix}
 1 ~&~ 0 ~&~ 0 \\
0 ~&~  0 ~&~ 1  \\
0 ~&~ 1 ~&~0
\end{pmatrix}
+g^{\nu}_{3}Y^{(4)}_{\mathbf{1'}}\begin{pmatrix}
0~&~ 0 ~&~ 1 \\
0 ~&~  1 ~&~ 0  \\
1 ~&~ 0 ~&~0
\end{pmatrix}
\right]\dfrac{v^2_u}{\Lambda}$}  \\
  &  &    \\ [-0.15in] \hline
$W_3$ & 3, ---  & {\scriptsize $   M_\nu = \left[g^{\nu}_{1}
\begin{pmatrix}
 2Y^{(6)}_{\mathbf{3}I,1} ~&~ -Y^{(6)}_{\mathbf{3}I,3} ~&~ -Y^{(6)}_{\mathbf{3}I,2} \\
-Y^{(6)}_{\mathbf{3}I,3} ~&~  2Y^{(6)}_{\mathbf{3}I,2} ~&~ -Y^{(6)}_{\mathbf{3}I,1}  \\
-Y^{(6)}_{\mathbf{3}I,2} ~&~ -Y^{(6)}_{\mathbf{3}I,1} ~&~2Y^{(6)}_{\mathbf{3}I,3}
\end{pmatrix}
+g^{\nu}_{2}\begin{pmatrix}
 2Y^{(6)}_{\mathbf{3}II,1} ~&~ -Y^{(6)}_{\mathbf{3}II,3} ~&~ -Y^{(6)}_{\mathbf{3}II,2} \\
-Y^{(6)}_{\mathbf{3}II,3} ~&~  2Y^{(6)}_{\mathbf{3}II,2} ~&~ -Y^{(6)}_{\mathbf{3}II,1}  \\
-Y^{(6)}_{\mathbf{3}II,2} ~&~ -Y^{(6)}_{\mathbf{3}II,1} ~&~2Y^{(6)}_{\mathbf{3}II,3}
\end{pmatrix}
+g^{\nu}_{3}Y^{(6)}_{\mathbf{1}}\begin{pmatrix}
1~&~ 0 ~&~ 0 \\
0 ~&~  0 ~&~ 1  \\
0 ~&~ 1 ~&~0
\end{pmatrix}
\right]\dfrac{v^2_u}{\Lambda} $} \\
   &  &     \\ [-0.15in]\hline
  &  &     \\ [-0.15in]
$S_1$ &  2, 0 &  {\scriptsize $ M_D = \begin{pmatrix}
2g^{D}_1Y^{(2)}_{\mathbf{3},1}        ~&~  (-g^{D}_1+g^{D}_2)Y^{(2)}_{\mathbf{3},3} ~&~ (-g^{D}_1-g^{D}_2)Y^{(2)}_{\mathbf{3},2} \\
(-g^{D}_1-g^{D}_2)Y^{(2)}_{\mathbf{3},3}  ~&~     2g^{D}_1Y^{(2)}_{\mathbf{3},2}    ~&~ (-g^{D}_1+g^{D}_2)Y^{(2)}_{\mathbf{3},1}  \\
 (-g^{D}_1+g^{D}_2)Y^{(2)}_{\mathbf{3},2} ~&~ (-g^{D}_1-g^{D}_2)Y^{(2)}_{\mathbf{3},1}  ~&~ 2g^{D}_1Y^{(2)}_{\mathbf{3},3}
\end{pmatrix}v_{u} $},~~{\scriptsize   $M_N = g^{M}\begin{pmatrix}
1 & 0 & 0 \\
0 & 0 & 1 \\
0 & 1 & 0 \\
\end{pmatrix}\Lambda $} \\
  &  &     \\ [-0.15in]\hline
  &  &    \\ [-0.15in]

$S_2$ & $-1$, 1  &  {\scriptsize $ M_D = g^{D}\begin{pmatrix}
1 & 0 & 0 \\
0 & 0 & 1 \\
0 & 1 &0 \\
\end{pmatrix}v_{u}$},~~{\scriptsize     $M_N = g^{M} \begin{pmatrix}
  2Y^{(2)}_{\mathbf{3},1} ~&~ -Y^{(2)}_{\mathbf{3},3} ~&~ -Y^{(2)}_{\mathbf{3},2} \\
 -Y^{(2)}_{\mathbf{3},3} ~&~ 2Y^{(2)}_{\mathbf{3},2}  ~&~ -Y^{(2)}_{\mathbf{3},1}  \\
 -Y^{(2)}_{\mathbf{3},2} ~&~ -Y^{(2)}_{\mathbf{3},1} ~&~2Y^{(2)}_{\mathbf{3},3} \\
\end{pmatrix} \Lambda$}  \\
 &  &     \\ [-0.15in] \hline

&  &     \\ [-0.15in]

$S_3$ & 1, 1  &   {\scriptsize $ M_D = \begin{pmatrix}
2g^{D}_1Y^{(2)}_{\mathbf{3},1}        ~&~  (-g^{D}_1+g^{D}_2)Y^{(2)}_{\mathbf{3},3} ~&~ (-g^{D}_1-g^{D}_2)Y^{(2)}_{\mathbf{3},2} \\
(-g^{D}_1-g^{D}_2)Y^{(2)}_{\mathbf{3},3}  ~&~     2g^{D}_1Y^{(2)}_{\mathbf{3},2}    ~&~ (-g^{D}_1+g^{D}_2)Y^{(2)}_{\mathbf{3},1}  \\
 (-g^{D}_1+g^{D}_2)Y^{(2)}_{\mathbf{3},2} ~&~ (-g^{D}_1-g^{D}_2)Y^{(2)}_{\mathbf{3},1}  ~&~ 2g^{D}_1Y^{(2)}_{\mathbf{3},3}
\end{pmatrix}v_{u} $},~~{\scriptsize$M_N = g^{M} \begin{pmatrix}
  2Y^{(2)}_{\mathbf{3},1} ~&~ -Y^{(2)}_{\mathbf{3},3} ~&~ -Y^{(2)}_{\mathbf{3},2} \\
 -Y^{(2)}_{\mathbf{3},3} ~&~ 2Y^{(2)}_{\mathbf{3},2}  ~&~ -Y^{(2)}_{\mathbf{3},1}  \\
 -Y^{(2)}_{\mathbf{3},2} ~&~ -Y^{(2)}_{\mathbf{3},1} ~&~2Y^{(2)}_{\mathbf{3},3} \\
\end{pmatrix}\Lambda$}   \\ \hline
   &   &     \\ [-0.15in]
$S_4$ & $-2,$ 2  & {\scriptsize $ M_D = g^{D}\begin{pmatrix}
1 & 0 & 0 \\
0 & 0 & 1 \\
0 & 1 &0 \\
\end{pmatrix}v_{u}$},~~{\scriptsize $   M_N = \left[g^{M}_{1}
\begin{pmatrix}
 2Y^{(4)}_{\mathbf{3},1} ~&~ -Y^{(4)}_{\mathbf{3},3} ~&~ -Y^{(4)}_{\mathbf{3},2} \\
-Y^{(4)}_{\mathbf{3},3} ~&~  2Y^{(4)}_{\mathbf{3},2} ~&~ -Y^{(4)}_{\mathbf{3},1}  \\
-Y^{(4)}_{\mathbf{3},2} ~&~ -Y^{(4)}_{\mathbf{3},1} ~&~2Y^{(4)}_{\mathbf{3},3}
\end{pmatrix}
+g^{M}_{2}Y^{(4)}_{\mathbf{1}}\begin{pmatrix}
 1 ~&~ 0 ~&~ 0 \\
0 ~&~  0 ~&~ 1  \\
0 ~&~ 1 ~&~0
\end{pmatrix}
+g^{M}_{3}Y^{(4)}_{\mathbf{1'}}\begin{pmatrix}
0~&~ 0 ~&~ 1 \\
0 ~&~  1 ~&~ 0  \\
1 ~&~ 0 ~&~0
\end{pmatrix}
\right]\Lambda$}  \\
  &  &    \\ [-0.15in] \hline
$S_5$ & $-3,$ 3  & {\scriptsize $ M_D = g^{D}\begin{pmatrix}
1 & 0 & 0 \\
0 & 0 & 1 \\
0 & 1 &0 \\
\end{pmatrix}v_{u}$},~~{\scriptsize $   M_N = \left[g^{M}_{1}
\begin{pmatrix}
 2Y^{(6)}_{\mathbf{3}I,1} ~&~ -Y^{(6)}_{\mathbf{3}I,3} ~&~ -Y^{(6)}_{\mathbf{3}I,2} \\
-Y^{(6)}_{\mathbf{3}I,3} ~&~  2Y^{(6)}_{\mathbf{3}I,2} ~&~ -Y^{(6)}_{\mathbf{3}I,1}  \\
-Y^{(6)}_{\mathbf{3}I,2} ~&~ -Y^{(6)}_{\mathbf{3}I,1} ~&~2Y^{(6)}_{\mathbf{3}I,3}
\end{pmatrix}
+g^{M}_{2}\begin{pmatrix}
 2Y^{(6)}_{\mathbf{3}II,1} ~&~ -Y^{(6)}_{\mathbf{3}II,3} ~&~ -Y^{(6)}_{\mathbf{3}II,2} \\
-Y^{(6)}_{\mathbf{3}II,3} ~&~  2Y^{(6)}_{\mathbf{3}II,2} ~&~ -Y^{(6)}_{\mathbf{3}II,1}  \\
-Y^{(6)}_{\mathbf{3}II,2} ~&~ -Y^{(6)}_{\mathbf{3}II,1} ~&~2Y^{(6)}_{\mathbf{3}II,3}
\end{pmatrix}
+g^{M}_{3}Y^{(6)}_{\mathbf{1}}\begin{pmatrix}
1~&~ 0 ~&~ 0 \\
0 ~&~  0 ~&~ 1  \\
0 ~&~ 1 ~&~0
\end{pmatrix}
\right]\Lambda $} \\
  \hline
  \hline
\end{tabular} }
\caption{\label{tab:neutrino}The predictions for the neutrino mass matrices, the neutrino masses are generated through the Weinberg operator for the models $W_{1,2,3}$ and the type-I seesaw mechanism for the models $S_{1,2,3,4,5}$. Here we only present the cases which involve at most three real coupling constants in the effective light neutrino mass matrix.}
\end{table}

\subsection{Phenomenologically viable lepton models}
\label{sec:lepton_models}

In section~\ref{sec:charged_lepton_sector} and section~\ref{sec:neutrino_sector}, we have systematically constructed the charged lepton and neutrino mass matrices respectively.
By diagonalizing the mass matrices we can obtain the charged lepton and neutrino masses and the lepton mixing matrix. The variation of the model parameters will dynamically affect the values of these experimental observables. In order to quantitatively estimate how well a model can describe the data, we perform a $\chi^2$ analysis to find out the best fit values of the input parameters and the corresponding predictions for lepton masses and mixing parameters. Because the inverted ordering (IO) neutrino mass spectrum is disfavored by the global data analyses~\cite{Esteban:2018azc, Capozzi:2017ipn} at about $3\sigma$ confidence level, we focus on normal ordering (NO) neutrino masses in this paper. From the NuFIT v5.0 with Super-Kamiokanda atmospheric data~\cite{Esteban:2020cvm},  the best fit values and $1\sigma$ ranges of the neutrino parameters are
\begin{equation}
\label{eq:mixing_data}
\begin{array}{c}
\sin ^{2} \theta_{12}=0.304_{-0.012}^{+0.012}\,, \quad \sin ^{2} \theta_{13}=0.02219_{-0.00063}^{+0.00062}\,, \quad \sin ^{2} \theta_{23}=0.573_{-0.020}^{+0.016}\,, \\
\delta_{C P}^{l} / \pi=1.0944_{-0.1333}^{+0.1500}\,, \quad \frac{\Delta m_{21}^{2}}{10^{-5} \mathrm{eV}^{2}}=7.42_{-0.20}^{+0.21}\,, \quad \frac{\Delta m_{31}^{2}}{10^{-3} \mathrm{eV}^{2}}=2.517_{-0.028}^{+0.026}\,,
\end{array}
\end{equation}
where $\theta_{12},\theta_{13}$ and $\theta_{23}$ are the three lepton mixing angles, $\delta_{CP}^{l}$ is the Dirac CP phase, $\Delta m_{21}^2$ and $\Delta m_{31}^2$ are the neutrino mass squared differences. The charge lepton masses will enter in the $\chi^2$ function in the form of their ratios, the best fit values and $1\sigma$ errors
are taken from Ref.~\cite{Antusch:2013jca},
\begin{equation}
\label{eq:mass_data}
m_{e}/m_{\mu}=0.004737 \pm 0.000040 , \quad m_{\mu}/m_{\tau}=0.05857 \pm 0.00047\,.
\end{equation}
at the grand unified theory (GUT) scale $2\times10^{16}$ GeV.
Notice that the charged lepton mass matrices are given at the scale where the modulus $\tau$ obtains the vacuum expectation value $\langle\tau\rangle$, which is expected to be around the GUT scale. The overall scale of the charged lepton mass matrix doesn't affect the mass ratios and mixing parameters, and it is fixed by the central value of $m_\tau=1.30234$~GeV at GUT scale. Similarly the overall scale of the light neutrino mass matrix is fixed by the solar neutrino mass squared difference $\Delta m_{21}^2=7.42\times 10^{-5}\text{eV}^2$. It has been shown that the effect of renormalization group evolution (RGE) on the neutrino masses and mixing parameters can be negligible for small value of $\tan\beta$ and NO neutrino masses, hence the corrections from RGE are not considered~\cite{Criado:2018thu}.

The minimization algorithm in \texttt{TMinuit}~\cite{minuit}, a package developed by CERN, is used to numerically minimize the value of $\chi^2$ function to determine the best fit values of the input parameters. The free parameters vary in some given ranges in \texttt{TMinuit}. We regard the complex modulus $\tau$ as a random complex number in the fundamental domain $\mathcal{F}: |\text{Re}\tau|\leq\frac{1}{2}$, $\text{Im}\tau>0$ and $|\tau|\geq1$. The absolute values of all coupling constants freely run in the range of $[0, 10^6]$.

Combining the possible structures in the charged lepton sector and neutrino sector, we can obtain $220\times8=1760$ lepton models.
In this paper, we are only interested in the minimal models that are compatible with the experimental data, thus only the lepton models with 7 real parameters~\textbf{(A)} and 8 real parameters~\textbf{(B)} are considered\footnote{The minimal $A_4$ modular models for leptons contain $6$ real parameters, nevertheless they are incompatible with the experimental data.}.  We perform a comprehensive numerical analysis for these models one by one.

\begin{table}[hptb!]
\centering
\begin{tabular}{c}
\begin{tabular}{|c|c|c||c|c|c|}  \hline\hline
  Models & $(k_{E_{1}^{c}}',k_{E_{2}^{c}}',k_{E_{3}^{c}}')$ & $(l_{1},l_{2},l_{3})$ &         Models & $(k_{E_{1}^{c}}',k_{E_{2}^{c}}',k_{E_{3}^{c}}')$ & $(l_{1},l_{2},l_{3})$ \\
  \hline
  $C_{1}$ & $(2,4,8)$ & $(0,0,1)$ & $C_{11}$ & $(2,2,6)$ & $(1,0,2)$ \\
  \hline
  $C_{2}$ & $(4,4,8)$ & $(1,0,1)$ & $C_{12}$ & $(2,4,6)$ & $(1,0,2)$\\
  \hline
  $C_{3}$ & $(4,4,8)$ & $(2,0,1)$ & $C_{13}$ & $(2,4,6)$ & $(1,1,2)$\\
  \hline
  $C_{4}$ & $(2,4,8)$ & $(1,0,1)$ & $C_{14}$ & $(2,2,6)$ & $(1,2,2)$\\
  \hline
  $C_{5}$ & $(2,4,8)$ & $(2,0,1)$ & $C_{15}$ & $(2,4,6)$ & $(1,2,2)$\\
  \hline
  $C_{6}$ & $(2,4,8)$ & $(0,1,2)$ & $C_{16}$ & $(2,2,6)$ & $(2,0,0)$\\
  \hline
  $C_{7}$ & $(4,4,8)$ & $(0,1,2)$ & $C_{17}$ & $(2,2,6)$ & $(2,1,0)$\\
  \hline
  $C_{8}$ & $(2,4,8)$ & $(1,1,2)$ & $C_{18}$ & $(2,4,6)$ & $(2,0,0)$\\
  \hline
  $C_{9}$ & $(4,4,8)$ & $(2,1,2)$ & $C_{19}$ & $(2,4,6)$ & $(2,1,0)$\\
  \hline
  $C_{10}$ & $(2,4,8)$ & $(2,1,2)$ & $C_{20}$ & $(2,4,6)$ & $(2,2,0)$\\
  \hline\hline
\end{tabular}
\end{tabular}
\caption{\label{tab:lepton_model_par7} List of the charged lepton models $C_{1}$, $C_2$,\ldots, $C_{20}$, where $k_{E_{i}^{c}}'\equiv k_{E_{i}^{c}}+k_L$ is the modular weight of the modular forms in the charged lepton Yukawa coupling. We would like to remind that the right-handed charged lepton $E^c_i$ transform as $\mathbf{1}^{l_{i}}$ under $A_4$ with $\mathbf{1}^{0}\equiv \mathbf{1}$, $\mathbf{1}^{1}\equiv\mathbf{1}'$ and $\mathbf{1}^{2}\equiv\mathbf{1}''$. }
\end{table}

\vskip 0.3cm
\noindent \textbf{(A) Models with $7$ real parameters}
\vskip 0.2cm

There are two possibilities: 3 couplings in the charged lepton sector and 2 couplings in the neutrino sector, or 4 couplings in the charged lepton sector and 1 coupling in the neutrino sector. The former case has $20\times2=40$ (number of charged lepton models times neutrino models) lepton models, and the latter case has $90\times2=180$ possible models.
We scan over all these models and optimize the $\chi^2$ function with the \texttt{TMinuit} package. We find that 20 lepton models out of them can give very good fit to the experimental data for certain values of input parameters. All the 20 models share the same neutrino model $S_2$ in table~\ref{tab:neutrino}, while the charged lepton models are different and they are listed in table~\ref{tab:lepton_model_par7}}, where the modular weights  $k_{E_{i}^{c}}'\equiv k_{E_{i}^{c}}+k_L$ and the representation indices $l_{i}$ with $i=1,2,3$ are given explicitly. The corresponding charged lepton mass matrices can be straightforwardly read out from table~\ref{Tab:R_{a}} with the given values of $k_{E_{i}^{c}}'$ and $l_{i}$. Furthermore we show the best fit values of the input parameters and the predictions of the mixing parameters and neutrino masses in table~\ref{tab:lepton_res_par7} for NO neutrino mass spectrum. It is notable that the first ten models $C_{1,\ldots,10}$-$S_2$ give quite similar predictions for the lepton observables at the best fit points, and the predictions of the other ten models $C_{11,\ldots,20}$-$S_2$ are also very similar. The models $C_{1,\ldots,5}$-$S_2$ only differ in the first row, the value of the coupling constant $\alpha_e$ associated with the first row should be smaller than $\beta_e$, $\gamma_{e,1}$ and $|\gamma_{e, 2}|$ in order to be accommodate the hierarchical charged lepton masses, as can be seen from table~\ref{tab:lepton_res_par7}. Numerically the contribution of the first row is found to be of order $10^{-3}$ with respect to the other two rows. Therefore the five models $C_{1,\ldots,5}$-$S_2$ give similar numerical results and the best fit values of $\tau$ are very close to each other. Analogously the models $C_{6,\ldots,10}$-$S_2$ are also only different in the first row whose contributions are negligible. Notice that the best fit values of $\tau$ of the models $C_{1,\ldots,5}$-$S_2$ and $C_{6,\ldots,10}$-$S_2$ are different, it is a numerical coincidence that they give similar predictions for the lepton masses and mixing parameters. Analogously the above reasoning also holds true for the latter ten models $C_{11,\ldots,20}$-$S_2$ except that the contribution of the second row rather than the first row is insignificant.
For illustration, we take the model $C_1$-$S_2$ as an example. We use the widely-used sampler \texttt{MultiNest}~\cite{Feroz:2007kg,Feroz:2008xx} to scan the parameter space fully and efficiently, and the predictions for the lepton masses and mixing parameters are required to be compatible with data at $3\sigma$ level. The correlations among the input parameters and observables are shown in figure~\ref{fig:C1S2}. We see that the CP violation phase $\delta^{l}_{CP}$ is predicted to be around $1.5\pi$ and the atmospheric mixing angle $\theta_{23}$ is in the second octant. These predictions could be tested in the forthcoming neutrino oscillation experiments.  

\begin{table}[t!]
\centering
\renewcommand{\arraystretch}{1.1}
\begin{tabular}{c}
\resizebox{1.0\textwidth}{!}{
\begin{tabular}{|c|c|c|c|c|c|c|c|c|c|c|}  \hline\hline
\multirow{2}{*}{ Models} & \multicolumn{7}{c|}{ Best fit values of the input parameters for NO} &\multirow{2}{*}{$\chi^2_{\mathrm{min}}$}  \\ \cline{2-8}
 & $\texttt{Re}\langle \tau \rangle$ &$\texttt{Im}\langle \tau \rangle$  & $\beta_{e}/\alpha_{e}$ & $\gamma_{e,1}/\alpha_{e}$  & $\gamma_{e,2}/\alpha_{e}$  & $\alpha_{e} v_d$/MeV  & $\dfrac{(g^{D})^2v_u^2}{g^{M}\Lambda}/$meV & \\ \hline
$C_1$-$S_2$ & 0.19214&1.09373&230.31529&1.20754$\times 10^{3}$&2.04179$\times 10^{3}$&0.32342&24.74650&4.87172 \\ \hline
$C_2$-$S_2$ & 0.19214&1.09373&183.94677&964.37095&1.63075$\times 10^{3}$&0.40495&24.74647&4.87191\\ \hline
$C_3$-$S_2$ & 0.19214&1.09373&62.75683&329.01303&556.35886&1.18694&24.74648&4.87191\\ \hline
$C_4$-$S_2$ & 0.19214&1.09373&132.87150&696.60007&1.17795$\times 10^{3}$&0.56061&24.74647&4.87191 \\ \hline
$C_5$-$S_2$ & 0.19214&1.09374&72.13887&378.32788&639.58609&1.03246&24.74642&4.86955 \\ \hline 
$C_6$-$S_2$ & $-$0.13677&1.18044&241.68127&3.75452$\times 10^{3}$&$-$2.35923$\times 10^{3}$&0.33938&23.58270&4.87172 \\ \hline
$C_7$-$S_2$ & $-0.13677$&1.18044&183.94677&2.85758$\times 10^{3}$&$-$1.79567$\times 10^{3}$&0.44590&23.58267&4.87191 \\ \hline
$C_8$-$S_2$ & $-$0.13678&1.18043&75.69807&1.17614$\times 10^{3}$&$-$739.00172&1.08340&23.58289&4.86955 \\ \hline
$C_9$-$S_2$ & $-$0.13677&1.18044&62.75684&974.91502&$-$612.62690&1.30698&23.58267&4.87191 \\ \hline
$C_{10}$-$S_2$ & $-$0.13677&1.18044&139.42871&2.166$\times 10^{3}$&$-$1.36109$\times 10^{3}$&0.58827&23.58267&4.87191 \\ \hline \hline
$C_{11}$-$S_2$ & 0.17243&1.14007&0.01051&11.53676&6.32548&67.00293&23.96081&9.61938 \\ \hline
$C_{12}$-$S_2$ & 0.17244&1.14004&0.00526&11.53706&6.32415&67.00313&23.96145&9.58973 \\ \hline
$C_{13}$-$S_2$ & 0.17235&1.14019&0.00504&11.53723&6.32610&67.00833&23.95954&9.59010 \\ \hline
$C_{14}$-$S_2$ & 0.17243&1.14007&0.01917&11.53674&6.32549&67.00294&23.96081&9.61938 \\ \hline
$C_{15}$-$S_2$ & 0.17248&1.13999&0.00503&11.53695&6.32352&67.00093&23.96217&9.59336 \\ \hline
$C_{16}$-$S_2$ & $-$0.17243&1.14007&0.01051&6.32548&11.53675&67.00297&23.96080&9.61938 \\ \hline
$C_{17}$-$S_2$ & $-$0.17243&1.14007&0.01917&6.32549&11.53674&67.00297&23.96080&9.61938 \\ \hline
$C_{18}$-$S_2$ & $-$0.17249&1.13996&0.00504&6.32449&11.53444&67.00001&23.96252&9.59010 \\ \hline
$C_{19}$-$S_2$ & $-$0.17241&1.14010&0.00526&6.32455&11.53774&67.00516&23.96072&9.58973 \\ \hline
$C_{20}$-$S_2$ & $-$0.17237&1.14015&0.00504&6.32468&11.53901&67.00708&23.95998&9.59336 \\ \hline\hline
\end{tabular}}\\[0.2in]
  \resizebox{1.0\textwidth}{!}{
  \begin{tabular}{|c|c|c|c|c|c|c|c|c|c|c|c|c|c|c|}  \hline
\multirow{2}{*}{ Models} & \multicolumn{13}{c|}{Predictions for mixing parameters and neutrino masses at best fit point}      \\ \cline{2-14}
& $\sin^2\theta_{12}$ &$\sin^2\theta_{13}$ &$\sin^2\theta_{23}$&$\delta^l_{CP}/\pi$ & $m_{e}/m_{\mu}$ & $m_{\mu}/m_{\tau}$ & $\Delta m^{2}_{21}/\Delta m^{2}_{31}$ & $\alpha_{21}/\pi$       &$\alpha_{31}/\pi$ & $m_1$/meV & $m_2$/meV & $m_3$/meV & $m_{\beta\beta}$/meV  \\ \hline
 $C_{1-10}$-$S_2$ & 0.30287&0.02218&0.58073&1.41681&0.00474&0.05857&0.02957&1.43461&0.92299&10.96085&13.94060&51.27982&8.57441 \\ \hline
$C_{11-20}$-$S_2$ & 0.33670&0.02162&0.57918&1.01474&0.00474&0.05857&0.03030&0.94333&1.85751&10.90683&13.89816&50.67471&3.52050\\
    \hline\hline
\end{tabular}}
\end{tabular}
\caption{\label{tab:lepton_res_par7}The best fit values of the input parameters at the minimum of the $\chi^2$ for NO neutrino masses. We give the predictions for neutrino mixing angles $\theta_{12}$, $\theta_{13}$, $\theta_{23}$ and Dirac CP violating phase $\delta^l_{CP}$ as well as Majorana CP violating phases $\alpha_{21}$, $\alpha_{31}$, and the light neutrino masses $m_{1,2,3}$ and the effective mass $m_{\beta\beta}$ in neutrinoless double beta decay.}
\end{table}

\vskip 0.3cm
\noindent \textbf{(B) Models with $8$ real parameters}
\vskip 0.2cm

This type of models involve 6 real coupling constants, and they can be decomposed into: 3 couplings in both neutrino and charged lepton sectors, or 4 couplings in charged lepton sector and 2 couplings in neutrino sector, or 5 couplings in charged lepton sector and 1 coupling in neutrino sector. There are totally 440 such models up to weight 8. In the same fashion as previous case, we numerically minimize the $\chi^2$ function with the \texttt{TMinuit} package. Eventually we find 360 phenomenologically viable lepton models for certain values of input parameters, and 206 models are consistent with the experimental data at $1\sigma$ level. Due to the limit of space, we present eight benchmark models, the structures of the charged lepton sector are given in table~\ref{tab:lepton_model_par8}. We show the best fit values of the input parameters and the predictions of the mixing parameters and neutrino masses in table~\ref{tab:lepton_res_par8} under the assumption of NO neutrino masses. The $A_4$ modular model with gCP symmetry was recently studied in~\cite{Okada:2020brs}, and a lepton model with 8 parameters was proposed. It corresponds to the $D_{2}$-$W_{2}$ model in our work, see table~\ref{tab:lepton_res_par8} for the numerical results of this model.

\begin{table}[hptb!]
\centering
\begin{tabular}{c}
\begin{tabular}{|c|c|c|}  \hline\hline
  Models & $(k_{E_{1}^{c}}',k_{E_{2}^{c}}',k_{E_{3}^{c}}')$ & $(l_{1},l_{2},l_{3})$ \\
  \hline
  $D_{1}$ & $(4,6,8)$ & $(1,2,2)$ \\
  \hline
  $D_{2}$ & $(2,2,2)$ & $(0,2,1)$ \\
  \hline
  $D_{3}$ & $(2,4,4)$ & $(0,1,2)$ \\
  \hline
  $D_{4}$ & $(4,4,6)$ & $(1,2,0)$ \\
  \hline
  $D_{5}$ & $(4,6,8)$ & $(0,1,1)$ \\
  \hline
  $D_{6}$ & $(2,2,8)$ & $(0,2,1)$ \\
  \hline
  $D_{7}$ & $(2,2,4)$ & $(0,1,0)$ \\
  \hline
$D_{8}$ & $(4,4,4)$ & $(0,1,2)$ \\
  \hline\hline
\end{tabular}
\end{tabular}
\caption{\label{tab:lepton_model_par8}List of the charged lepton models $D_{1}$, $D_2$,\ldots, $D_{8}$, where $k_{E_{i}^{c}}'\equiv k_{E_{i}^{c}}+k_L$ is the weight of the modular forms in the charged lepton Yukawa coupling. The left-handed lepton fields are assumed to be an $A_4$ triplet and the right-handed changed leptons $E^c_i$ transform as $\mathbf{1}^{l_{i}}$ under $A_4$.}
\end{table}

\begin{table}[ht!]
\centering
\renewcommand{\arraystretch}{1.1}
\begin{tabular}{c}
\resizebox{1.0\textwidth}{!}{
    \begin{tabular}{|c|c|c|c|c|c|c|}
      \hline\hline
\multicolumn{7}{|c|}{Best fit values of the input parameters for NO}\\ \hline
Model & $D_{1}-W_{1}$  & Model & $D_{4}-S_{1}$ & Model & $D_{2}-W_{2}$ & $D_{3}-W_{3}$  \\\hline
$\texttt{Re}(\tau)$ &  0.05389  & $\texttt{Re}(\tau)$ &0.35597 &$\texttt{Re}(\tau)$ &  0.06026 &0.01908  \\
$\texttt{Im}(\tau)$ &  2.66993 & $\texttt{Im}(\tau)$ &0.94804 & $\texttt{Im}(\tau)$ & 1.01109 &1.02906 \\
$\beta_{e,1}/\alpha_{e}$ &  2.29253$\times 10^{3}$ & $\beta_{e}/\alpha_{e}$ &2.78535$\times 10^{3}$ & $\beta_{e}/\alpha_{e}$ & 51.55285   &15.57674  \\
$\beta_{e,2}/\alpha_{e}$ & 4.22108$\times 10^{4}$ & $\gamma_{e,1}/\alpha_{e}$ &558.22251 & $\gamma_{e}/\alpha_{e}$ & 765.43156 &0.03590 \\
$\gamma_{e,1}/\alpha_{e}$ & 1.21762$\times 10^{3}$ & $\gamma_{e,2}/\alpha_{e}$ &$-$651.31064 & $g^{\nu}_{2}/g^{\nu}_{1}$ &  14.03290 &$-$0.30519 \\
$\gamma_{e,2}/\alpha_{e}$ & 2.91946$\times 10^{4}$ & $g^{D}_{2}/g^{D}_{1}$ &3.54628 & $g^{\nu}_{3}/g^{\nu}_{1}$ &  1.14817 &7.97382 \\
$\alpha_{e} v_d/\text{MeV}$ &  3.35046 &$\alpha v_d/\text{MeV}$ &0.26571 &$\alpha v_d/\text{MeV}$ &  1.32754 &60.61083  \\
$(v_u^2/\Lambda)/\text{meV}$ & 28.48955 &  $((g^{D}_{1})^{2}v_u^2/(g^{M}\Lambda))/\text{meV}$ &2.69366 & $(g^{\nu}_{1}v_u^2/\Lambda)/\text{meV}$ & 3.51922 &12.59531 \\\hline \hline
Model & $D_{5}-S_{2}$ & Model & $D_{6}-S_{3}$ & Model & $D_{7}-S_{4}$ & $D_{8}-S_{5}$ \\\hline
$\texttt{Re}(\tau)$ &  $-$0.17648 & $\texttt{Re}(\tau)$ &$-$0.08357 & $\texttt{Re}(\tau)$ & 0.23585&  $-$0.30537 \\
$\texttt{Im}(\tau)$ &  1.12772 & $\texttt{Im}(\tau)$ & 1.15143 & $\texttt{Im}(\tau)$ & 1.47319& 1.77322  \\
$\beta_{e,1}/\alpha_{e}$ &   69.56803 & $\beta_{e}/\alpha_{e}$ &120.79494 & $\beta_{e}/\alpha_{e}$ &16.52692& 3.53453$\times 10^{3}$ \\
$\beta_{e,2}/\alpha_{e}$ & 158.37007 & $\gamma_{e,1}/\alpha_{e}$ & 20.66057 & $\gamma_{e}/\alpha_{e}$ & 0.01649& 209.08250  \\
$\gamma_{e,1}/\alpha_{e}$ & 3.93725$\times 10^{3}$ & $\gamma_{e,2}/\alpha_{e}$ & $-$22.46201 & $g^{M}_{2}/g^{M}_{1}$ & $-$2.78133& $-$2.66807  \\
$\gamma_{e,2}/\alpha_{e}$ & $-$1.61631$\times 10^{3}$ & $g^{D}_{2}/g^{D}_{1}$ & $-$0.76278 & $g^{M}_{3}/g^{M}_{1}$ & 5.54678& $-$0.07778  \\
$\alpha v_d/\text{MeV}$ & 0.33971 &  $\alpha v_d/\text{MeV}$ &4.99601 & $\alpha v_d/\text{MeV}$ &  75.92891& 0.36424 \\
$((g^{D})^{2}v_u^2/(g^{M}\Lambda))/\text{meV}$ & 24.23812 &  $((g^{D}_{1})^{2}v_u^2/(g^{M}\Lambda))/\text{meV}$ &16.74812 & $((g^{D})^{2}v_u^2/(g^{M}_{1}\Lambda))/\text{meV}$ & 56.40935& 26.95432 \\\hline
   \end{tabular}
}\\[8.4em]
  \hskip -0.09cm
  \resizebox{0.9945\textwidth}{!}{
\begin{tabular}{|c|c|c|c|c|c|c|c|c|c|c|c|c|c|c|} \hline
\multirow{2}{*}{ Models} & \multicolumn{13}{c|}{Predictions for mixing parameters and neutrino masses at best fitting point}  \\ \cline{2-14}
& $\sin^2\theta_{12}$ &$\sin^2\theta_{13}$ &$\sin^2\theta_{23}$&$\delta^l_{CP}/\pi$ & $m_{e}/m_{\mu}$ & $m_{\mu}/m_{\tau}$ & $\Delta m^{2}_{21}/\Delta m^{2}_{31}$ & $\alpha_{21}/\pi$  &$\alpha_{31}/\pi$ & $m_1$/meV & $m_2$/meV & $m_3$/meV & $m_{\beta\beta}$/meV  \\ \hline
$D_{1}$-$W_1$ & 0.33989&0.02244&0.53179&1.11139&0.00474&0.05853&0.02989&1.06921&0.14655&27.84586&29.14261&56.98847&10.04516 \\ \hline
  $D_{2}$-$W_2$ & 0.30933&0.02242&0.52707&1.09701&0.00474&0.05857&0.02906&0.00931&1.00870&63.29939&63.88046&80.93137&60.55211 \\ \hline
    $D_{3}$-$W_3$ & 0.30895&0.02234&0.55798&1.24021&0.00474&0.05857&0.02917&0.14358&1.41214&18.10917&20.04600&53.49454&16.76825\\ \hline
    $D_{4}$-$S_1$ & 0.30993&0.02237&0.56318&1.24722&0.00474&0.05857&0.02923&1.30471&0.49605&27.07896&28.41074&57.10709&16.19580\\ \hline
    $D_{5}$-$S_2$ & 0.31005&0.02237&0.56302&1.33116&0.00474&0.05857&0.02924&0.14241&1.52824&10.96356&13.93197&51.45661&10.40884\\ \hline
    $D_{6}$-$S_3$ & 0.31244&0.02235&0.56370&1.15976&0.00474&0.05857&0.02925&1.01310&0.35970&2.57262&8.97320&50.32765&0.11038\\ \hline
    $D_{7}$-$S_4$ & 0.30972&0.02237&0.56394&1.19850&0.00474&0.05857&0.02924&1.05455&1.94805&9.32428&12.68236&51.13357&3.22886\\ \hline
$D_{8}$-$S_5$ & 0.31123&0.02230&0.56132&1.47812&0.00474&0.05857&0.02921&0.05237&0.82489&12.08684&14.83212&51.72936&13.65139 \\    \hline\hline
\end{tabular}}
\end{tabular}
\caption{\label{tab:lepton_res_par8}The best fit values of the input parameters at the minimum of the $\chi^2$ under the assumption of NO neutrino masses. We give the predictions for neutrino mixing angles $\theta_{12}$, $\theta_{13}$, $\theta_{23}$, and Dirac CP violating phase $\delta^l_{CP}$ as well as Majorana CP violating phases $\alpha_{21}$, $\alpha_{31}$, and the light neutrino masses $m_{1,2,3}$ and the effective mass $m_{\beta\beta}$ in neutrinoless double beta decay.
}
\end{table}

\begin{figure}[hptb!]
\centering
\includegraphics[width=6.5in]{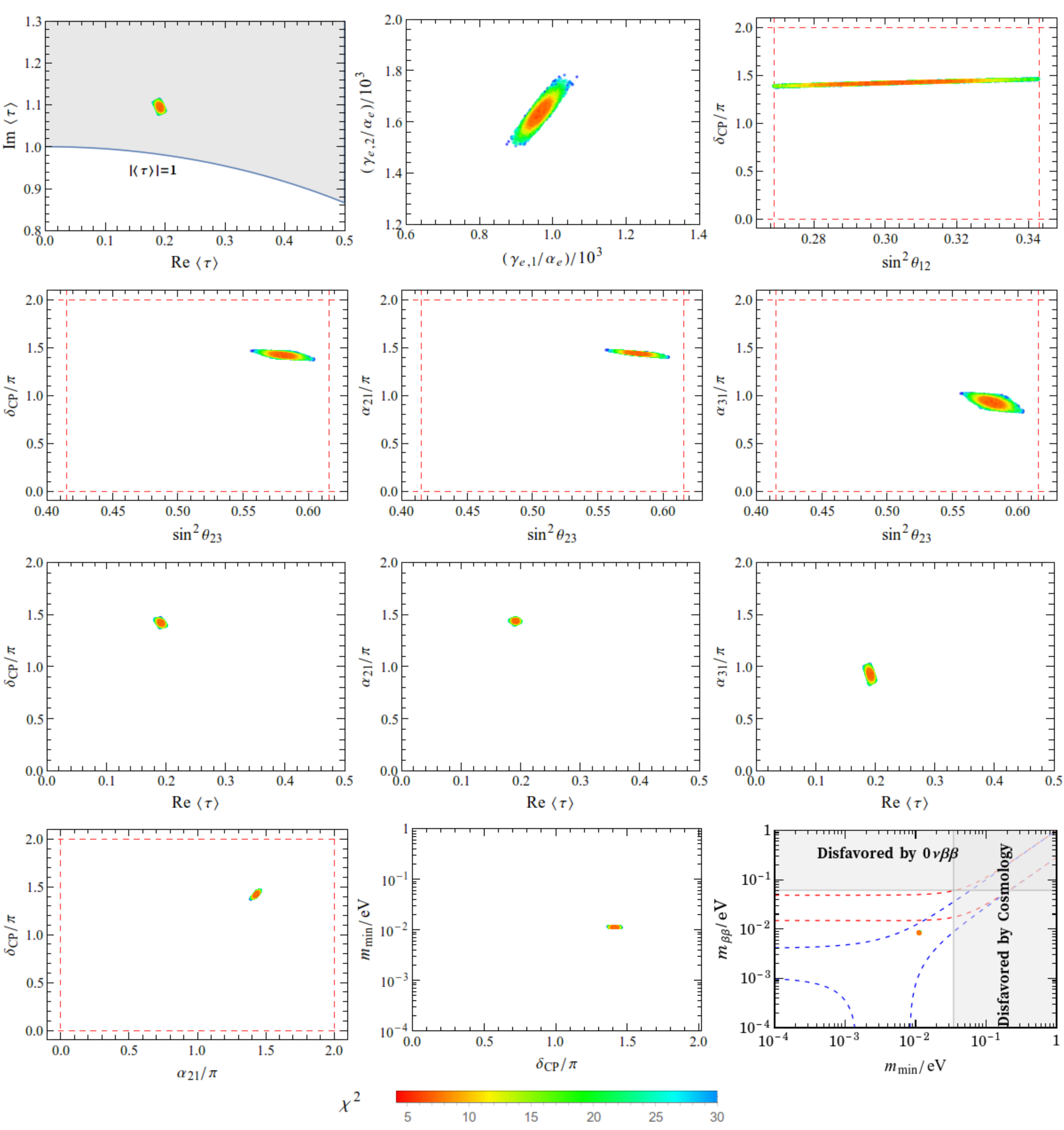}
\caption{The predictions for the correlations among the input free parameters, neutrino mixing angles, CP violation phases and neutrino masses in the lepton model $C_1$-$S_2$. The vertical and horizontal dashed lines are the $3\sigma$ bounds taken from~\cite{Esteban:2020cvm}. }
\label{fig:C1S2}
\end{figure}

\section{Quark models based on $A_{4}$ modular symmetry with gCP}
\label{sec:quark_sector}

In this section, we proceed to consider the quark models in the framework of $A_{4}$ modular symmetry with gCP. Similar to what we have done for the charged lepton sector in section~\ref{sec:charged_lepton_sector}, we can systematically analyze possible quark models with $A_{4}$ modular symmetry. The three generations of left-handed quark fields are assigned to a triplet $\mathbf{3}$ of the $A_{4}$ modular group, while the three generations of right-handed quarks are singlets of $A_{4}$, i.e.,
\begin{equation}
  Q_{L}\equiv (Q_{1},Q_{2},Q_{3})^{T}\sim \mathbf{3}\,,\quad q_{i}^{c}\sim \mathbf{1}^{l_{i}}~~\text{with}~~q=u,d,~~i=1,2,3,
\end{equation}
where the convention for $\mathbf{1}^{l_{i}}$ is the same as that in Eq.~\eqref{eq:W_E}. The modular weights of left- and right-handed quarks are denoted as $k_{Q_{L}}$ and $k_{q^{c}_{i}}$ respectively. The superpotential for the quark masses is determined by the assignments of right-handed quarks and the values of $k_{Q_{L}}+k_{q^{c}_{i}}$,
\begin{equation}
\label{eq:quark_yukawa}
  \mathcal{W}_{q}=\sum_{i=1}^{3}\sum_{a}\alpha_{i,a}q_{i}^{c}Q_{L}H_{u/d}Y_{\mathbf{3}a}^{(k_{Q_{L}}+k_{q^{c}_{i}})}\,,
\end{equation}
where $\alpha_{i,a}$ are coupling constants. The general form of the $(ij)$ entry of the quark mass matrix $M_{q}$ is
\begin{equation}
\label{eq:general-ele-quark} (M_{q})_{ij}=\sum_{a}\alpha_{i,a}v_{u/d}Y_{\mathbf{3}a,3-\text{mod}(l_{i}+j+1,3)}^{(k_{Q_{L}}+k_{q^{c}_{i}})}\,.
\end{equation}
Analogous to the charged lepton sector in section~\ref{sec:charged_lepton_sector}, if we are only concerned with modular forms up to weight $8$, then the explicit form of $R^q_i$ ($i$-th row of $M_{q}$) with $k_{Q_{L}}+k_{q^{c}_{i}}=2,4,6,8$ can be obtained from table~\ref{Tab:R_{a}} with the replacements $k_{L}+k_{E^{c}_{i}}\rightarrow k_{Q_{L}}+k_{q^{c}_{i}}$. As a result, the quark mass matrix can also take $220$ possible structures. According to the number of free coupling parameters in quark mass matrix,
we can divide the 220 possible quark mass matrices into four distinct classes: $M_{Q}^{I}$, $M_{Q}^{II}$, $M_{Q}^{III}$ and $M_{Q}^{IV}$ in which 3, 4, 5 and 6 couplings are introduced in the quark mass matrix respectively. The number of models are $20$, $90$, $90$ and $20$ respectively for these four classes.

As shown in above discussion, both up and down quark mass matrices can take $220$ possible structures with proper weight and representation assignments of quark fields. In a concrete model where both up and down quark mass matrices are involved, the possible number of real couplings will range from $6$ to $12$ besides the complex modulus $\tau$.
There are totally $10$ observables in quark sector, including six quark masses $m_{u,c,t}$, $m_{d,s,b}$, three quark mixing angles $\theta_{12}^{q}$, $\theta_{13}^{q}$, $\theta_{23}^{q}$, and one quark $CP$ violation phase $\delta_{CP}^{q}$. The quark model will not be attractive if too many free parameters are required to fit the ten quark observables. Hence we only focus on the models with number of free parameters no more than $11$.
More specifically, we only consider the following four kinds of quark models.

\begin{itemize}
\item[\textbf{(I)}]{\bf{Quark model with 8 free real parameters including $\texttt{Re}(\tau)$ and $\texttt{Im}(\tau)$}}

In this case, there are total $6$ free couplings in the up and down quark mass matrices, which can only originated from the assignment $M_{u}\in M_{Q}^{I}$ and $M_{d}\in M_{Q}^{I}$. The triplet modular forms that enter the quark superpotentials in Eq.~\eqref{eq:quark_yukawa} can only be $Y_{\mathbf{3}}^{(2)}$ and $Y_{\mathbf{3}}^{(4)}$. Combining the up and down quark mass matrices, we can obtain $20\times 20=400$ models which have $8$ free real parameters.

\item[\textbf{(II)}]{\bf{Quark model with 9 free real parameters including $\texttt{Re}(\tau)$ and $\texttt{Im}(\tau)$}}

For the quark models with 9 free parameters, the classifications of quark mass matrices should be
  \begin{equation}
    M_{u}\in M_{Q}^{I}\,,\quad M_{d} \in M_{Q}^{II}\,,\quad \text{or} \quad M_{u}\in M_{Q}^{II}\,,\quad M_{d} \in M_{Q}^{I}\,.
  \end{equation}
There are $20\times 90\times 2=3600$ distinct models with 9 free parameters.

\item[\textbf{(III)}]{\bf{Quark model with 10 free real parameters including $\texttt{Re}(\tau)$ and $\texttt{Im}(\tau)$}}

  There are three types of the combinations of the up and down quark mass matrices in this case,
  \begin{eqnarray} \nonumber
    &&M_{u}\in M_{Q}^{I}\,,\quad M_{d} \in M_{Q}^{III}\,,\\ \nonumber
    \text{or} \quad &&M_{u}\in M_{Q}^{III}\,,\quad M_{d} \in M_{Q}^{I}\,,\\
    \text{or} \quad &&M_{u}\in M_{Q}^{II}\,,\quad M_{d} \in M_{Q}^{II}\,.
  \end{eqnarray}
So there are total $20\times 90\times 2+90\times 90=11700$ models with $10$ free real parameters.

\item[\textbf{(IV)}]{\bf{Quark model with 11 free real parameters including $\texttt{Re}(\tau)$ and $\texttt{Im}(\tau)$}}

If there are $11$ parameters in quark mass matrices, we find that they should satisfy the following conditions,
  \begin{eqnarray} \nonumber
    &&M_{u}\in M_{Q}^{I}\,,\quad M_{d} \in M_{Q}^{IV}\,,\\ \nonumber
    \text{or} \quad &&M_{u}\in M_{Q}^{IV}\,,\quad M_{d} \in M_{Q}^{I}\,,\\ \nonumber
    \text{or} \quad &&M_{u}\in M_{Q}^{II}\,,\quad M_{d} \in M_{Q}^{III}\,,\\
    \text{or} \quad &&M_{u}\in M_{Q}^{III}\,,\quad M_{d} \in M_{Q}^{II}\,.
  \end{eqnarray}
  Counting all the possibilities, we can obtain $2\times 20\times 20+2\times 90\times 90=17000$ models with $11$ free real parameters.
\end{itemize}

Similar to the lepton sector, we need to find the phenomenologically viable quark models through the $\chi^2$ analysis, the definition of the $\chi^2$ function involves experimental data of the ratios of quark masses and quark mixing parameters,
\begin{eqnarray}
\label{eq:quark_data}
\nonumber&&m_{u} / m_{c}=(1.9286 \pm 0.6017) \times 10^{-3}\,, \quad m_{c} / m_{t}=(2.8213 \pm 0.1195) \times 10^{-3}\,,\\
\nonumber&&m_{d} / m_{s}=(5.0523 \pm 0.6191) \times 10^{-2}\,, \quad m_{s} / m_{b}=(1.8241 \pm 0.1005) \times 10^{-2}\,, \\
\nonumber&&\theta_{12}^{q}=0.22736 \pm 0.00073\,, \quad \theta_{13}^{q}=0.00349 \pm 0.00013\,,\\
         && \theta_{23}^{q}=0.04015 \pm 0.00064\,,\quad \delta_{C P}^{q}/\pi=0.3845 \pm 0.0173\,.
\end{eqnarray}
The central values and $1\sigma$ ranges at GUT scale are taken from Ref.~\cite{Antusch:2013jca} with $\tan\beta=10$ and the SUSY breaking scale $M_{\text{SUSY}}=10$~TeV. Analogous to the lepton sector, we use the central value of $m_{t}=87.4555 \mathrm{GeV}$ and $m_{b}=0.9682 \mathrm{GeV}$ at GUT scale to determine the overall coefficients of the quark mass matrices $M_u$ and $M_d$ respectively.

We scan the parameter space of the above quark models one by one with the \texttt{TMinuit} package.
Unfortunately, we have not found a model with 8 parameters that is consistent with the experimental data, and the minimum of $\chi^2$ is very large.
There is also no model with 9 parameters which can accommodate the experimental data at $3\sigma$ confidence level. But some models with 9 parameters can achieve a relatively small $\chi^2$, and only the mixing angle $\theta_{23}^q$ marginally lies outside the $3\sigma$ allowed region. It's reasonable to regard these models as a good leading order approximation, we will give such an example below. For models with 10 or 11 parameters,
we find thousands of models can accommodate the experimental data. It is too lengthy to show all these phenomenologically viable models, consequently  only some examples are provided in this work and they are given in section~\ref{sec:unified_models} within the complete models for leptons and quarks.

Here we present an example of the quark models with 9 parameters. In this model, the quark fields transform under modular symmetry $A_{4}$ as follows,
\begin{eqnarray}\nonumber
  &&Q_{L}\sim \mathbf{3}\,,~~u^{c}\sim \mathbf{1}\,,~~c^{c}\sim \mathbf{1'}\,,~~t^{c}\sim \mathbf{1''}\,,~~d^{c}\sim \mathbf{1''}\,,~~s^{c}\sim \mathbf{1'}\,,~~b^{c}\sim \mathbf{1''}\,,\\
  && k_{Q_{L}}=2-k_{u^{c}}=2-k_{c^{c}}=2-k_{t^{c}}=2-k_{d^{c}}=4-k_{s^{c}}=6-k_{b^{c}}\,,
\end{eqnarray}
where the modular weight $k_{Q_{L}}$ is a general integer.
The modualr invariant superpotentials are given as
\begin{eqnarray}\nonumber
  \mathcal{W}_{u}&=&\alpha_{u}u^{c}_{\mathbf{1}}(Q_{L}Y^{(2)}_{\mathbf{3}})_{\mathbf{1}}H_{u} + \beta_{u}c^{c}_{\mathbf{1'}}(Q_{L}Y^{(2)}_{\mathbf{3}})_{\mathbf{1''}}H_{u} + \gamma_{u}t^{c}_{\mathbf{1''}}(Q_{L}Y^{(2)}_{\mathbf{3'}})_{\mathbf{1'}}H_{u}\,,\\
  \nonumber
  \mathcal{W}_{d}&=&\alpha_{d}d^{c}_{\mathbf{1''}}(Q_{L}Y^{(2)}_{\mathbf{3}})_{\mathbf{1'}}H_{d} + \beta_{d}s^{c}_{\mathbf{1'}}(Q_{L}Y^{(4)}_{\mathbf{3}})_{\mathbf{1''}}H_{d} + \gamma_{d,1}b^{c}_{\mathbf{1''}}(Q_{L}Y^{(6)}_{\mathbf{3}I})_{\mathbf{1'}}H_{d}\\
  &~&  + \gamma_{d,2}b^{c}_{\mathbf{1''}}(Q_{L}Y^{(6)}_{\mathbf{3}II})_{\mathbf{1'}}H_{d}\,.
\end{eqnarray}
The corresponding up and down quark mass matrices can be written as
\begin{eqnarray} \nonumber
    M_{u}&=&\left( \begin{array}{ccc}
                   \alpha_{u}Y^{(2)}_{\mathbf{3},1}~ & \alpha_{u}Y^{(2)}_{\mathbf{3},3} & ~ \alpha_{u}Y^{(2)}_{\mathbf{3},2}  \\
                   \beta_{u}Y^{(2)}_{\mathbf{3},3}~ & \beta_{u}Y^{(2)}_{\mathbf{3},2} & ~ \beta_{u}Y^{(2)}_{\mathbf{3},1}  \\
                   \gamma_{u}Y^{(2)}_{\mathbf{3},2}~ & \gamma_{u}Y^{(2)}_{\mathbf{3},1} & ~ \gamma_{u}Y^{(2)}_{\mathbf{3},3} \end{array} \right)v_u \,,\\
   M_{d}&=&\left( \begin{array}{ccc}
                   \alpha_{d}Y^{(2)}_{\mathbf{3},2} & \alpha_{d}Y^{(2)}_{\mathbf{3},1} & \alpha_{d}Y^{(2)}_{\mathbf{3},3}  \\
                   \beta_{d}Y^{(4)}_{\mathbf{3},1} & \beta_{d}Y^{(4)}_{\mathbf{3},3} & \beta_{d}Y^{(4)}_{\mathbf{3},2}  \\
                   \gamma_{d,1}Y^{(6)}_{\mathbf{3}I,2}+\gamma_{d,2}Y^{(6)}_{\mathbf{3}II,2} ~ & \gamma_{d,1}Y^{(6)}_{\mathbf{3}I,1}+\gamma_{d,2}Y^{(6)}_{\mathbf{3}II,1} & ~ \gamma_{d,1}Y^{(6)}_{\mathbf{3}I,3}+\gamma_{d,2}Y^{(6)}_{\mathbf{3}II,3}  \end{array} \right)v_d \,.
\end{eqnarray}
The best fit values of the input parameters are given as follows
\begin{equation}
\begin{gathered}
\langle\tau\rangle=0.49175+0.88563i\,,\quad \beta_{u}/\alpha_{u}=518.22933\,,\quad \gamma_u/\alpha_{u}=1.83596\times 10^{5}\,, \\
\beta_{d}/\alpha_{d}=9.39751\,,\quad \gamma_{d,1}/\alpha_{d}=32.46046\,,\quad \gamma_{d,2}/\alpha_{d}=-0.02697\,,\\
\alpha_{u} v_u=0.00034~\text{GeV},\quad \alpha_d v_d=0.05081~\text{GeV}\,.
\end{gathered}
\end{equation}
The quark mixing parameters and mass ratios are determined to be
\begin{equation}
\begin{gathered}
\theta^q_{12}=0.22734\,,\quad \theta^q_{13}=0.00332\,,\quad \theta^q_{23}=0.05708\,,\quad \delta^q_{CP}=0.39532 ~\pi\,,\\
m_u/m_c=0.00193\,,~ m_c/m_t=0.00282\,,~ m_d/m_s=0.05055\,,~ m_s/m_b=0.01815\,.
\end{gathered}
\end{equation}
We find that only $\theta_{23}^{q}$ is somewhat large, all other observables are within the $3\sigma$ ranges of the experimental data given in Eq.~\eqref{eq:quark_data}. Notice that if we adopt the experimental data of quark masses and mixings used in Ref.~\cite{Okada:2020rjb}, this model would be compatible with data very well\footnote{In Ref.~\cite{Okada:2020rjb}, the central values as well the $1\sigma$ errors of the quark masses and mixing parameters are taken from~\cite{Bjorkeroth:2015ora}.
Comparing the experimental data used in Ref.~\cite{Okada:2020rjb} with those in the present work, we can find that the best-fit values of quark Yukawa couplings and the CKM mixing parameters are almost the same while the $1\sigma$ ranges given in~\cite{Okada:2020rjb} are larger than the corresponding ones used in our paper. To be more specific, the $1\sigma$ erros of $\theta^q_{12}$ and $\theta^q_{23}$ in~\cite{Okada:2020rjb} are about $5$ and $10$ times as large as the corresponding values in this paper respectively, while the $1\sigma$ regions of other observables are about twice as large as those in Eq.~\eqref{eq:quark_data}.}.

\section{Unified models of Leptons and Quarks}
\label{sec:unified_models}

In the above two sections, we have discussed the phenomenologically viable models in lepton sector and quark sector separately. Now we shall investigate whether the $A_4$ modular can describe the flavor structures of quark and lepton simultaneously, where the complex modulus $\tau$ works as a portal to combine these two sectors.
All coupling constants are real because of the gCP symmetry, thus the vacuum expectation value of $\tau$ is the only source of CP violation phases of both quarks and leptons.

In section~\ref{sec:lepton_sector}, we have found 20 lepton models with 7 parameters and 360 lepton models with 8 parameters that can accommodate the experimental data. In section~\ref{sec:quark_sector}, we have found thousands of quark models with 10 or 11 parameters that can explain the experimental data of quark masses and mixing. In each sector, the experimentally $3\sigma$-allowed $\tau$s are recorded in the \texttt{TMinuit} optimization processes. By showing the $\tau$ sample of lepton sector and quark sector in the same plot, we can see whether there are overlapping regions. In this way, we can easily determine whether
the experimental data of quarks and leptons can be described in a given combination of lepton and quark models. It turns out that the lepton models with 7 parameters ($L_7$) together with the quark models with 10 parameters ($Q_{10}$) failed to accommodate the experimental data, while the combinations of $L_7$-$Q_{11}$ and $L_8$-$Q_{10}$ can explain the measured masses and mixings of both quarks and leptons.
We can find thousands of these two kinds of unified models. Due to the limitation of the length of the article, we only give one example of $L_7$-$Q_{11}$ unification and one example of $L_8$-$Q_{10}$ unification in the following.

\subsection{Unified model of $L_7$-$Q_{11}$}
\label{sec:unified_7_11}

\begin{figure}[hptb!]
\centering
\includegraphics[width=6.5in]{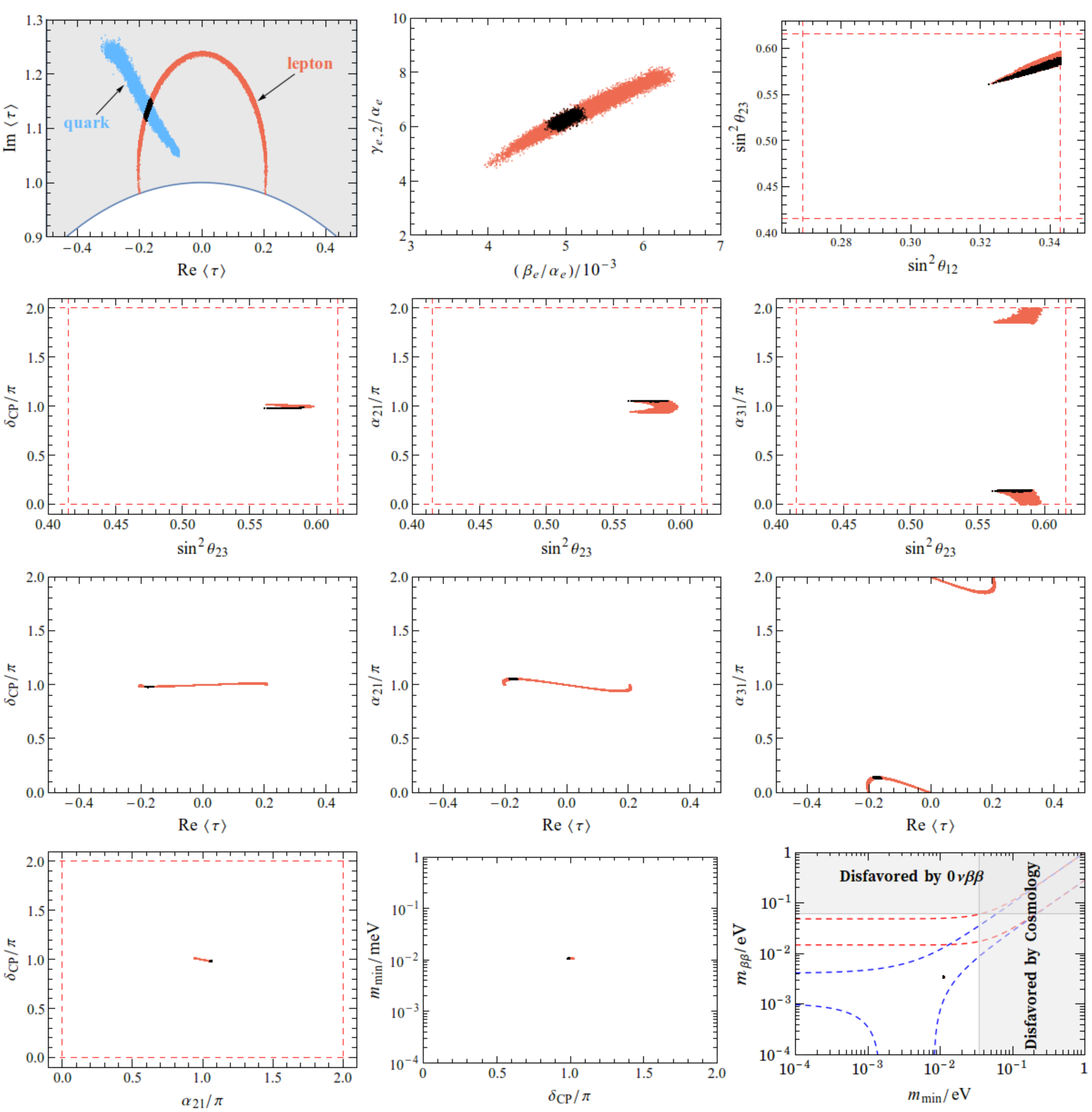}
\caption{The predictions for the correlations among the input free parameters, neutrino mixing angles, CP violation phases and neutrino masses in the unified model $L_7$-$Q_{11}$. The red and blue areas denote points compatible with experimental data of quarks and leptons respectively, and the black is the overlapping region. The vertical and horizontal dashed lines are the $3\sigma$ bounds taken from~\cite{Esteban:2020cvm}.
}
\label{fig:L7Q11}
\end{figure}

In this section, we give an example of the unified model with 7 parameters in the lepton sector and 11 parameters in the quark sector.
The complex modulus $\tau$ in the quark and lepton sectors is the same one, so the total number of real free parameters is $7+11-2=16$. Thus this model is very predictive, and it use 16 real free parameters to describe the 22 masses and mixing parameters of quarks and leptons. In the lepton sector, the model is denoted as $C_{13}$-$S_2$, which has been described in section~\ref{sec:lepton_models}. For the sake of clarity, we list the details of the model in the following. The representation assignments and the modular weights of the lepton fields are,
  \begin{eqnarray}\nonumber
    &&L\sim \mathbf{3}\,,~~e^{c}\sim \mathbf{1'}\,,~~\mu^{c}\sim \mathbf{1'}\,,~~\tau^{c}\sim \mathbf{1''}\,,~~N^{c}\sim \mathbf{3}\,,\\
    &&k_{L}=-1\,,~~k_{e^{c}}=3\,,~~k_{\mu^{c}}=5\,,~~k_{\tau^{c}}=7\,\bl{,}~~k_{N^{c}}=1\,.
  \end{eqnarray}
The corresponding modular invariant lepton superpotentials are given by
\begin{eqnarray}\nonumber
  \mathcal{W}_{E}&=&\alpha_{e} e^{c}_{\mathbf{1'}}(LY^{(2)}_{\mathbf{3}})_{\mathbf{1''}}H_{d} + \beta_{e} \mu^{c}_{\mathbf{1'}}(LY^{(4)}_{\mathbf{3}})_{\mathbf{1''}}H_{d} + \gamma_{e,1} \tau^{c}_{\mathbf{1''}}(LY^{(6)}_{\mathbf{3}I})_{\mathbf{1'}}H_{d}+\gamma_{e,2} \tau^{c}_{\mathbf{1''}}(LY^{(6)}_{\mathbf{3}II})_{\mathbf{1'}}H_{d}\,,\\
  \mathcal{W}_{\nu}&=&g^{D}H_{u}(N^{c}L)_{\mathbf{1}}+ g^{M}\Lambda \left( (N^{c}N^{c})_{\mathbf{3}_{S}}Y^{(2)}_{\mathbf{3}}\right)_{\mathbf{1}}\,.
\end{eqnarray}
The corresponding charged lepton and neutrino mass matrices read as
\begin{eqnarray} \nonumber
  M_{E}&=&\left( \begin{array}{ccc}
                   \alpha_{e} Y^{(2)}_{\mathbf{3},3} ~& \alpha_{e} Y^{(2)}_{\mathbf{3},2} & ~\alpha_{e} Y^{(2)}_{\mathbf{3},1}  \\
                   \beta_{e} Y^{(4)}_{\mathbf{3},3} ~& \beta_{e} Y^{(4)}_{\mathbf{3},2} & ~\beta_{e} Y^{(4)}_{\mathbf{3},1}  \\
                  \gamma_{e,1} Y^{(6)}_{\mathbf{3}I,2}+\gamma_{e,2} Y^{(6)}_{\mathbf{3}II,2} ~& \gamma_{e,1} Y^{(6)}_{\mathbf{3}I,1} + \gamma_{e,2} Y^{(6)}_{\mathbf{3}II,1} & ~\gamma_{e,1} Y^{(6)}_{\mathbf{3}I,3} + \gamma_{e,2} Y^{(6)}_{\mathbf{3}II,3}  \\
                 \end{array} \right) v_{d}\,,\\
  M_D &=& g^{D}\begin{pmatrix}
1 & ~0~ & 0 \\
0 & ~0~ & 1 \\
0 & ~1~ &0 \\
\end{pmatrix}v_{u} \,,\quad
  M_N =g^{M} \begin{pmatrix}
  2Y^{(2)}_{\mathbf{3},1} ~&~ -Y^{(2)}_{\mathbf{3},3} ~&~ -Y^{(2)}_{\mathbf{3},2} \\
 -Y^{(2)}_{\mathbf{3},3} ~&~ 2Y^{(2)}_{\mathbf{3},2}  ~&~ -Y^{(2)}_{\mathbf{3},1}  \\
 -Y^{(2)}_{\mathbf{3},2} ~&~ -Y^{(2)}_{\mathbf{3},1} ~&~2Y^{(2)}_{\mathbf{3},3} \\
\end{pmatrix} \Lambda\,.
\end{eqnarray}
The quark fields transform under the $A_4$ modular symmetry as follow,
\begin{eqnarray}\nonumber
  &&Q_{L}\sim \mathbf{3}\,,~~u^{c}\sim \mathbf{1}\,,~~c^{c}\sim \mathbf{1}\,,~~t^{c}\sim \mathbf{1}\,,~~d^{c}\sim \mathbf{1}\,,~~s^{c}\sim \mathbf{1'}\,,~~b^{c}\sim \mathbf{1'}\,,\\
  && k_{Q_{L}}=2-k_{u^{c}}=4-k_{c^{c}}=6-k_{t^{c}}=4-k_{d^{c}}=6-k_{s^{c}}=8-k_{b^{c}}\,.
\end{eqnarray}
Then the Yukawa superpotentials for the quark masses are
\begin{eqnarray}\nonumber
  \mathcal{W}_{u}&=&\alpha_{u}u^{c}_{\mathbf{1}}(Q_{L}Y^{(2)}_{\mathbf{3}})_{\mathbf{1}}H_{u} + \beta_{u}c^{c}_{\mathbf{1}}(Q_{L}Y^{(4)}_{\mathbf{3}})_{\mathbf{1}}H_{u} + \gamma_{u,1}t^{c}_{\mathbf{1}}(Q_{L}Y^{(6)}_{\mathbf{3}I})_{\mathbf{1}}H_{u}\\ \nonumber
                 &~& + \gamma_{u,2}t^{c}_{\mathbf{1}}(Q_{L}Y^{(6)}_{\mathbf{3}II})_{\mathbf{1}}H_{u}\,,\\
  \nonumber
  \mathcal{W}_{d}&=&\alpha_{d}d^{c}_{\mathbf{1}}(Q_{L}Y^{(4)}_{\mathbf{3}})_{\mathbf{1}}H_{d} + \beta_{d,1}s^{c}_{\mathbf{1'}}(Q_{L}Y^{(6)}_{\mathbf{3}I})_{\mathbf{1''}}H_{d} + \beta_{d,2}s^{c}_{\mathbf{1'}}(Q_{L}Y^{(6)}_{\mathbf{3}II})_{\mathbf{1''}}H_{d}\\
  &~& + \gamma_{d,1}b^{c}_{\mathbf{1'}}(Q_{L}Y^{(8)}_{\mathbf{3}I})_{\mathbf{1''}}H_{d} + \gamma_{d,2}b^{c}_{\mathbf{1'}}(Q_{L}Y^{(8)}_{\mathbf{3}II})_{\mathbf{1''}}H_{d}\,.
\end{eqnarray}
We find the up and down quark mass matrices take the following form
\begin{eqnarray}\nonumber
    M_{u}&=&\left( \begin{array}{ccc}
                   \alpha_{u}Y^{(2)}_{\mathbf{3},1} ~& \alpha_{u}Y^{(2)}_{\mathbf{3},3} &~ \alpha_{u}Y^{(2)}_{\mathbf{3},2}  \\
                   \beta_{u}Y^{(4)}_{\mathbf{3},1} ~& \beta_{u}Y^{(4)}_{\mathbf{3},3} &~ \beta_{u}Y^{(4)}_{\mathbf{3},2}  \\
                   \gamma_{u,1}Y^{(6)}_{\mathbf{3}I,1}+\gamma_{u,2}Y^{(6)}_{\mathbf{3}II,1} ~& \gamma_{u,1}Y^{(6)}_{\mathbf{3}I,3}+\gamma_{u,2}Y^{(3)}_{\mathbf{3}II,3} &~ \gamma_{u,1}Y^{(6)}_{\mathbf{3}I,2}+\gamma_{u,2}Y^{(3)}_{\mathbf{3}II,2}  \\ \end{array} \right)v_{u} \,,\\
   M_{d}&=&\left( \begin{array}{ccc}
                   \alpha_{d}Y^{(4)}_{\mathbf{3},1} ~& \alpha_{d}Y^{(4)}_{\mathbf{3},3} &~ \alpha_{d}Y^{(4)}_{\mathbf{3},2}  \\
                   \beta_{d,1}Y^{(6)}_{\mathbf{3}I,3}+\beta_{d,2}Y^{(6)}_{\mathbf{3}II,3} ~& \beta_{d,1}Y^{(6)}_{\mathbf{3}I,2}+\beta_{d,2}Y^{(6)}_{\mathbf{3}II,2} &~ \beta_{d,1}Y^{(6)}_{\mathbf{3}I,1}+\beta_{d,2}Y^{(6)}_{\mathbf{3}II,1}  \\
                   \gamma_{d,1}Y^{(8)}_{\mathbf{3}I,3}+\gamma_{d,2}Y^{(8)}_{\mathbf{3}II,3} ~& \gamma_{d,1}Y^{(8)}_{\mathbf{3}I,2}+\gamma_{d,2}Y^{(8)}_{\mathbf{3}II,2} &~ \gamma_{d,1}Y^{(8)}_{\mathbf{3}I,1}+\gamma_{d,2}Y^{(8)}_{\mathbf{3}II,1}  \\ \end{array} \right)v_{d} \,.
\end{eqnarray}
We perform a global fit to the experimental data of quark and lepton simultaneously, the best fit values of the input parameters are determined to be
\begin{equation}
\begin{gathered}
 \langle\tau\rangle=-0.17492 + 1.13648i\,,\quad \beta_{e}/\alpha_{e} =0.00500\,,\quad \gamma_{e,1}/\alpha_{e}=11.54025\,, \\
  \gamma_{e,2}/\alpha_{e}=6.27248\,,\quad \alpha_{e}v_{d} = 0.06688~\text{MeV}\,,\quad \frac{(g^{D})^{2}v_{u}^{2}}{g^{M}\Lambda} = 23.94439~\text{meV}\,,\\
 \beta_{u}/\alpha_{u}=462.725\,,\quad \gamma_{u,1}/\alpha_{u}=7.66928\times 10^{3}\,,\quad \gamma_{u,2}/\alpha_{u}=-1.51202\times 10^{5}\,,\\
 \beta_{d,1}/\alpha_{d}=235.8\,,\quad \beta_{d,2}/\alpha_{d}=14.197\,,\quad \gamma_{d,1}/\alpha_{d}=94.9129\,, \\
 \quad \gamma_{d,2}/\alpha_{d}= 3.84761 \,,\quad \alpha_{u} v_u=0.00046 ~\text{GeV},\quad \alpha_d v_d=0.00298~\text{GeV}\,.
\end{gathered}
\end{equation}
The corresponding predictions for masses and mixing parameters of leptons and quarks are given as
\begin{eqnarray}
\nonumber&& \sin\theta^l_{12}=0.33632\,,\quad \sin\theta^l_{13}=0.02164\,,\quad \sin\theta^l_{23}=0.57884\,,\quad \delta^l_{CP}=0.98516\pi\,,\\
\nonumber&& \alpha_{21}=1.05680\pi\,,\quad \alpha_{31}=0.14266\pi\,,\quad m_e/m_{\mu}=0.00474\,,\quad m_{\mu}/m_{\tau}=0.05857\,,\\
\nonumber&&  \frac{\Delta m_{21}^{2}}{\Delta m_{31}^{2}} = 0.03040\,,\quad  m_1=10.87722~\text{meV}\,,\quad m_2=13.86412~\text{meV}\,,\\
\nonumber&&   m_3=50.48811~\text{meV}\,,\quad \sum_i m_i = 75.22945~\text{meV}\,,\quad m_{\beta\beta}=3.51842~\text{meV}\,,\\
\nonumber&& \theta^q_{12}=0.22719\,,\quad \theta^q_{13}=0.00350\,,\quad \theta^q_{23}=0.03988\,,\quad \delta^q_{CP}=0.38850~\pi \,,\\
         && m_u / m_c=0.00166\,,\quad m_c / m_t=0.00289\,,\quad m_d/m_s=0.04605\,,\quad m_s / m_b=0.01837\,.
\end{eqnarray}
We find that all of these observables are predicted to lie in the experimentally preferred $3\sigma$ ranges. We use the package \texttt{Multinest} to comprehensively scan the parameter space of lepton sector and quark sector independently, we require all the observables are in the experimentally preferred $3\sigma$ regions, and the results are plotted in figure~\ref{fig:L7Q11}. We can see there is indeed an overlapping region in the $\tau$ plane, which is colored in black.
It is obvious that the allowed value of $\tau$ shrink to a small region if we include the experimental constraints of both quarks and leptons. The allowed values of all observables scatter in narrow ranges around the best fit values.

\subsection{Unified model of $L_8$-$Q_{10}$}
\label{sec:unified_8_10}

\begin{figure}[hptb!]
\centering
\includegraphics[width=6.5in]{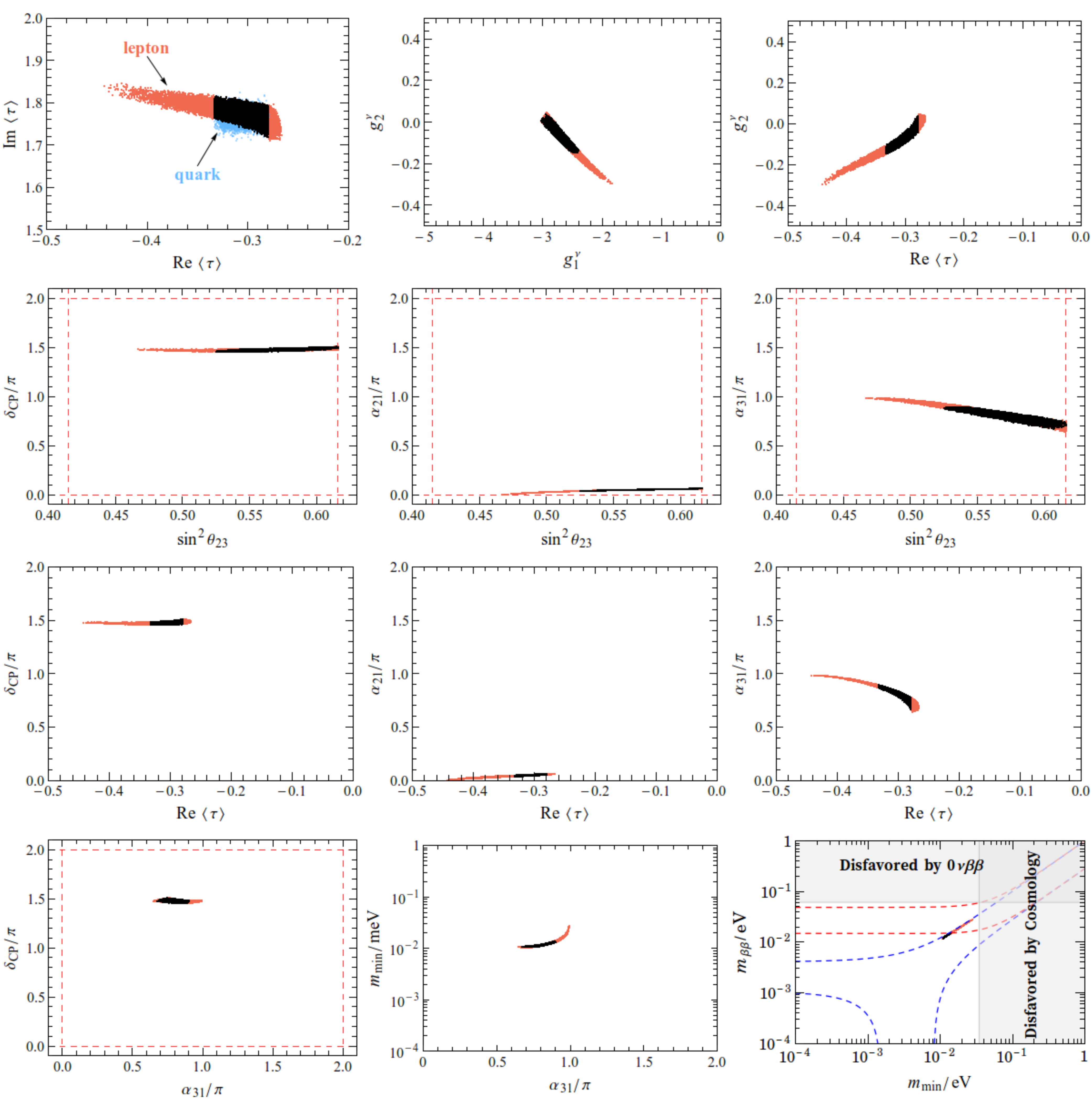}
\caption{The predictions for the correlations among the input free parameters, neutrino mixing angles, CP violation phases and neutrino masses of the unified model $L_8$-$Q_{10}$. We adopt the same convention as figure~\ref{fig:L7Q11}.
}
\label{fig:L8Q10}
\end{figure}

In this section, we give another example of the unified model with 8 parameters in the lepton sector and 10 parameters in the quark sector. The total number of real free parameters in this unified model is also $8+10-2=16$. The lepton model is $D_8$-$S_5$, and the transformation properties of the lepton fields are,
  \begin{eqnarray}\nonumber
    &&L\sim \mathbf{3}\,,~~e^{c}\sim \mathbf{1}\,,~~\mu^{c}\sim \mathbf{1'}\,,~~\tau^{c}\sim \mathbf{1''}\,,~~N^{c}\sim \mathbf{3}\,,\\
    &&k_{L}=-3\,,~~k_{e^{c}}=7\,,~~k_{\mu^{c}}=7\,,~~k_{\tau^{c}}=7\,\bl{,}~~k_{N^{c}}=3\,.
  \end{eqnarray}
Then we can read out the superpotentials of charged leptons and neutrinos as
\begin{eqnarray} \nonumber
  \mathcal{W}_{E}&=&\alpha_{e} e^{c}_{\mathbf{1}}(LY^{(4)}_{\mathbf{3}})_{\mathbf{1}}H_{d} + \beta_{e} \mu^{c}_{\mathbf{1'}}(LY^{(4)}_{\mathbf{3}})_{\mathbf{1''}}H_{d} + \gamma_{e} \tau^{c}_{\mathbf{1''}}(LY^{(4)}_{\mathbf{3}})_{\mathbf{1'}}H_{d}\,,\\ \nonumber
  \mathcal{W}_{\nu}&=&g^{D}H_{u}(N^{c}L)_{\mathbf{1}}+ g^{M}_{1}\Lambda \left( (N^{c}N^{c})_{\mathbf{3}_{S}}Y^{(6)}_{\mathbf{3}I}\right)_{\mathbf{1}}+ g^{M}_{2}\Lambda \left( (N^{c}N^{c})_{\mathbf{3}_{S}}Y^{(6)}_{\mathbf{3}II}\right)_{\mathbf{1}}\\
  &~&+ g^{M}_{3}\Lambda (N^{c}N^{c})_{\mathbf{1}}Y^{(6)}_{\mathbf{1}}\,,
\end{eqnarray}
which gives rise to
\begin{eqnarray} \nonumber
  M_{E}&=&\left( \begin{array}{ccc}
                   \alpha_{e} Y^{(4)}_{\mathbf{3},1} ~& \alpha_{e} Y^{(4)}_{\mathbf{3},3} &~ \alpha_{e} Y^{(4)}_{\mathbf{3},2}  \\
                   \beta_{e} Y^{(4)}_{\mathbf{3},3} ~& \beta_{e} Y^{(4)}_{\mathbf{3},2} &~ \beta_{e} Y^{(4)}_{\mathbf{3},1}  \\
                  \gamma_{e} Y^{(4)}_{\mathbf{3},2} ~& \gamma_{e} Y^{(4)}_{\mathbf{3},1} &~ \gamma_{e} Y^{(4)}_{\mathbf{3},3}  \\
                 \end{array} \right) v_{d}\,,\quad  M_D = g^{D}\begin{pmatrix}
1 ~&~ 0 ~&~ 0 \\
0 ~&~ 0 ~&~ 1 \\
0 ~&~ 1 ~&~ 0 \\
\end{pmatrix}v_{u}\,,\\ \nonumber
  M_N &=& \Bigg[ g^{M}_{1}
\begin{pmatrix}
 2Y^{(6)}_{\mathbf{3}I,1} ~&~ -Y^{(6)}_{\mathbf{3}I,3} ~&~ -Y^{(6)}_{\mathbf{3}I,2} \\
-Y^{(6)}_{\mathbf{3}I,3} ~&~  2Y^{(6)}_{\mathbf{3}I,2} ~&~ -Y^{(6)}_{\mathbf{3}I,1}  \\
-Y^{(6)}_{\mathbf{3}I,2} ~&~ -Y^{(6)}_{\mathbf{3}I,1} ~&~2Y^{(6)}_{\mathbf{3}I,3}
\end{pmatrix}+g^{M}_{2}\begin{pmatrix}
 2Y^{(6)}_{\mathbf{3}II,1} ~&~ -Y^{(6)}_{\mathbf{3}II,3} ~&~ -Y^{(6)}_{\mathbf{3}II,2} \\
-Y^{(6)}_{\mathbf{3}II,3} ~&~  2Y^{(6)}_{\mathbf{3}II,2} ~&~ -Y^{(6)}_{\mathbf{3}II,1}  \\
-Y^{(6)}_{\mathbf{3}II,2} ~&~ -Y^{(6)}_{\mathbf{3}II,1} ~&~2Y^{(6)}_{\mathbf{3}II,3}
\end{pmatrix}\\
&&
+g^{M}_{3}Y^{(6)}_{\mathbf{1}}\begin{pmatrix}
1~&~ 0 ~&~ 0 \\
0 ~&~  0 ~&~ 1  \\
0 ~&~ 1 ~&~0
\end{pmatrix}
\Bigg]\Lambda\,.
\end{eqnarray}
The assignments for the quark fields are
\begin{eqnarray}\nonumber
  &&Q_{L}\sim \mathbf{3}\,,~~u^{c}\sim \mathbf{1}\,,~~c^{c}\sim \mathbf{1}\,,~~t^{c}\sim \mathbf{1}\,,~~d^{c}\sim \mathbf{1}\,,~~s^{c}\sim \mathbf{1'}\,,~~b^{c}\sim \mathbf{1''}\,,\\
  && k_{Q_{L}}=2-k_{u^{c}}=4-k_{c^{c}}=6-k_{t^{c}}=2-k_{d^{c}}=2-k_{s^{c}}=8-k_{b^{c}}\,.
\end{eqnarray}
The modular invariant superpotentials for quark masses read as follows,
\begin{eqnarray}\nonumber
  \mathcal{W}_{u}&=&\alpha_{u}u^{c}_{\mathbf{1}}(Q_{L}Y^{(2)}_{\mathbf{3}})_{\mathbf{1}}H_{u} + \beta_{u}c^{c}_{\mathbf{1}}(Q_{L}Y^{(4)}_{\mathbf{3}})_{\mathbf{1}}H_{u} + \gamma_{u,1}t^{c}_{\mathbf{1}}(Q_{L}Y^{(6)}_{\mathbf{3}I})_{\mathbf{1}}H_{u}\\ \nonumber
&~& + \gamma_{u,2}\left(t^{c}_{\mathbf{1}}(Q_{L}Y^{(6)}_{\mathbf{3}II})_{\mathbf{1}}\right)_{\mathbf{1}}H_{u}\,,\\
  \nonumber
  \mathcal{W}_{d}&=&\alpha_{d}d^{c}_{\mathbf{1}}(Q_{L}Y^{(2)}_{\mathbf{3}})_{\mathbf{1}}H_{d} + \beta_{d,1}s^{c}_{\mathbf{1'}}(Q_{L}Y^{(2)}_{\mathbf{3}})_{\mathbf{1''}}H_{d} + \gamma_{d,1}b^{c}_{\mathbf{1''}}(Q_{L}Y^{(8)}_{\mathbf{3}I})_{\mathbf{1'}}H_{d}\\
  &~&  + \gamma_{d,2}b^{c}_{\mathbf{1''}}(Q_{L}Y^{(8)}_{\mathbf{3}II})_{\mathbf{1'}}H_{d}\,.
\end{eqnarray}
We find the up and down quark mass matrices are given by
  \begin{eqnarray} \nonumber
    M_{u}&=&\left( \begin{array}{ccc}
                   \alpha_{u}Y^{(2)}_{\mathbf{3},1} ~& \alpha_{u}Y^{(2)}_{\mathbf{3},3} &~ \alpha_{u}Y^{(2)}_{\mathbf{3},2}  \\
                   \beta_{u}Y^{(4)}_{\mathbf{3},1} ~& \beta_{u}Y^{(4)}_{\mathbf{3},3} &~ \beta_{u}Y^{(4)}_{\mathbf{3},2}  \\
                   \gamma_{u,1}Y^{(6)}_{\mathbf{3}I,1}+\gamma_{u,2}Y^{(6)}_{\mathbf{3}II,1} ~& \gamma_{u,1}Y^{(6)}_{\mathbf{3}I,3}+\gamma_{u,2}Y^{(6)}_{\mathbf{3}II,3} & ~\gamma_{u,1}Y^{(6)}_{\mathbf{3}I,2}+\gamma_{u,2}Y^{(6)}_{\mathbf{3}II,2}  \\ \end{array} \right)v_u \,,\\
   M_{d}&=&\left( \begin{array}{ccc}
                   \alpha_{d}Y^{(2)}_{\mathbf{3},1} ~& \alpha_{d}Y^{(2)}_{\mathbf{3},3} &~ \alpha_{d}Y^{(2)}_{\mathbf{3},2}  \\
                   \beta_{d}Y^{(2)}_{\mathbf{3},3} ~& \beta_{d}Y^{(2)}_{\mathbf{3},2} &~ \beta_{d}Y^{(2)}_{\mathbf{3},1}  \\
                   \gamma_{d,1}Y^{(8)}_{\mathbf{3}I,2}+\gamma_{d,2}Y^{(8)}_{\mathbf{3}II,2} ~& \gamma_{d,1}Y^{(8)}_{\mathbf{3}I,1}+\gamma_{d,2}Y^{(8)}_{\mathbf{3}II,1} &~ \gamma_{d,1}Y^{(8)}_{\mathbf{3}I,3}+\gamma_{d,2}Y^{(8)}_{\mathbf{3}II,3}  \\ \end{array} \right)v_d \,.
  \end{eqnarray}
The agreement between the model predictions and the experimental data is optimized for the following values of the input parameters are
\begin{equation}
\begin{gathered}
  \langle\tau\rangle=-0.30537 + 1.77322i\,,\quad \beta_{e}/\alpha_{e} =3.53453\times 10^{3}\,,\quad \gamma_{e}/\alpha_{e}=209.08250\,, \\
  g^{M}_{2}/g^{M}_{1}=-2.66807 \,,\quad  g^{M}_{3}/g^{M}_{1} = -0.07778 \,, \quad \alpha_{e}v_{d} = 0.36424 ~\text{MeV}\,,\\
  \frac{(g^{D})^{2}v_{u}^{2}}{g^{M}_{1}\Lambda} = 26.95432~\text{meV}\,,\quad \beta_{u}/\alpha_{u}=43.07455\,,\quad \gamma_{u,1}/\alpha_{u}=634.51584\,,\\
 \quad \gamma_{u,2}/\alpha_{u}=5.16218\times 10^{4}\,,\quad  \beta_{d}/\alpha_{d}=232.29418\,,\quad  \gamma_{d,1}/\alpha_{d}=0.99699\,, \\ \gamma_{d,2}/\alpha_{d}= -14.39314 \,,\quad
 \alpha_{u} v_u=0.00569 ~\text{GeV},\quad \alpha_d v_d=0.00412~\text{GeV}\,.
\end{gathered}
\end{equation}
The values of the masses and mixing parameters of leptons and quarks at the above best fit point are determined to be
\begin{eqnarray}
\nonumber&& \sin\theta^l_{12}=0.31123\,,\quad \sin\theta^l_{13}=0.02230\,,\quad \sin\theta^l_{23}=0.56132\,,\quad \delta^l_{CP}=1.47812\pi\,,\\
\nonumber&& \alpha_{21}=0.05237\pi\,,\quad \alpha_{31}=0.82489\pi\,,\quad m_e/m_{\mu}=0.00474\,,\quad m_{\mu}/m_{\tau}=0.05857\,,\\
\nonumber&&  \frac{\delta m_{21}^{2}}{\delta m_{31}^{2}} = 0.02921\,,\quad  m_1=12.08684~\text{meV}\,,\quad m_2=14.83212~\text{meV}\,,\\
\nonumber&&   m_3=51.72936~\text{meV}\,,\quad \sum_i m_i = 89.64832~\text{meV}\,,\quad m_{\beta\beta}=13.65139~\text{meV}\,,\\
\nonumber&& \theta^q_{12}=0.22736\,,\quad \theta^q_{13}=0.00348\,,\quad \theta^q_{23}=0.04119\,,\quad \delta^q_{CP}=0.38606~\pi \,,\\
&& m_u / m_c=0.00193\,,\quad m_c / m_t=0.00282\,,\quad m_d/m_s=0.05057\,,\quad m_s / m_b=0.01824\,,
\end{eqnarray}
which are in the experimentally favored $3\sigma$ regions. In figure~\ref{fig:L8Q10}, we show the correlations among the input free parameters, neutrino masses and mixing parameters predicted in the unified model $L_8$-$Q_{10}$. It is notable that the lepton Dirac CP phase $\delta^{l}_{CP}$ is close to maximal violation value $1.5\pi$ in this model.

\section{Conclusion}
\label{sec:conclusion}

In this paper, we have performed a systematical analysis of $\Gamma_{3}\cong A_{4}$ modular flavor models for leptons and quarks with gCP. We intend to understand how the minimal modular symmetry $A_4$ can help to understand the flavor structure of quarks and leptons, and we aim to find out the minimal models. It has been established that the CP transformation consistent with modular symmetry acts on the modulus as $\tau\rightarrow -\tau^{*}$. The representation matrices of the generators $S$ and $T$ are unitary and symmetric in our working basis, therefore the CP symmetry is exactly the canonical one and all the coupling constants are constrained to be real. In this setup, both modular symmetry and gCP are spontaneously broken by the vacuum expectation value of the complex modulus $\tau$. In particular, all weak CP violation phases arises from the real part of $\tau$.

The left-handed charged fermions are assumed to transform as triplet of $A_{4}$, and the right-handed charged fermions are assigned to singlets under $A_{4}$. Here the charged fermions refer to charged leptons, up-type quarks and down-type quarks. We find out the most general form of charged lepton and quark mass matrices given by Eq.~\eqref{eq:general-ele-lepton} and Eq.~\eqref{eq:general-ele-quark} respectively.
The three generations of right-handed charged fermions are distinguished from each other by the weight and representation assignments, then we can obtain 220 possible structures of charged fermion mass matrices if the weights of involved modular forms are less than 9. Moreover, we consider the Weinberg operator and the type-I seesaw mechanism to generate neutrino masses. In type-I seesaw mechanism, the right-handed neutrinos are assigned to be a $A_4$ triplet as well. The most general form of neutrino mass matrix is given in Eqs.~(\ref{eq:general-ele-neutrino-Wein},\ref{eq:general-ele-neutrino-SS}).
Requiring the number of coupling constants in the effective light neutrino mass matrix is less than 4, we can obtain $3$ Weinberg operator models $W_{1,2,3}$ and $5$ seesaw models $S_{1,2,3,4,5}$, they are summarized in table~\ref{tab:neutrino}.
We perform a $\chi^{2}$ analysis to estimate the goodness-of-fit of a $A_4$ modular model to the data. We find the minimal phenomenologically viable models for leptons have 7 real free parameters including the real and imaginary parts of $\tau$. The best fit values of the input parameters and the corresponding predictions for neutrino masses and mixing parameters are given in table~\ref{tab:lepton_res_par7}. For the next-to-minimal models with $8$ real input parameters, we find $360$ models which can explain the observed masses and mixings of leptons, and the numerical results of some typical models are presented in table~\ref{tab:lepton_res_par8}.

In quark sector, we have systematically classified the $A_4$ modular quark models based on the number of free parameters. All the quark models with 8, 9, 10 and 11 input parameters are discussed. There are no viable quark models
with $8$ or $9$ parameters.
However, we find thousands of quark models which can accommodate the experimental data of quark masses and CKM mixing matrix with 10 or 11 parameters. They can combine with the lepton models to give a unified description of both quarks and leptons. We give two predictive examples of quark-lepton unification models in which the 22 masses and mixing parameters of quarks and leptons are described in terms of 14 free real coupling constants and a common complex modulus $\tau$.

Because the weights and representations of the matter fields are not subject to any constraint in the modular invariance approach, many possible models could be constructed, as explicit shown for the $A_4$ modular symmetry. We expect this drawback could possibly be overcame in the appealing framework of eclectic flavor groups which is a maximal extension of the traditional flavor group by finite discrete modular symmetries~\cite{Nilles:2020nnc,Nilles:2020kgo}. The scheme of eclectic flavor symmetry can severely
restrict the possible group representations and modular weights of matter fields. Thus the structure of both K\"ahler potential and superpotential is under control. It would be much desirable that only few models in the bottom-up approach would survive in view of the eclectic flavor symmetry.
In summary, the finite modular group $A_{4}$ provides a simple and economical framework to understand the flavor structures of the quarks and leptons.

\section*{Acknowledgements}

JNL and GJD are supported by the National Natural Science Foundation of China under Grant Nos 11975224,  11835013 and 11947301. CYY is supported in part by the Grants No.~NSFC-11975130, No.~NSFC-12035008, No.~NSFC-12047533, by the National Key Research and Development Program of China under Grant No. 2017YFA0402200 and the China Post-doctoral Science Foundation under Grant No. 2018M641621.

\clearpage
\newpage


\providecommand{\href}[2]{#2}\begingroup\raggedright\endgroup

\end{document}